\documentclass[aps,pra,twocolumn,superscriptaddress,showpacs,nofootinbibfloatfix,amsmath,amsfonts,amssymb]{revtex4-2}%
\usepackage{amsmath,amsfonts,amssymb,color}
\usepackage{amsthm}
\usepackage{leftidx}
\usepackage{graphicx}
\usepackage{xcolor}
\usepackage{dcolumn}
\usepackage{bm}
\usepackage{epstopdf}
\usepackage{epsfig}
\usepackage{environ}
\usepackage{pdfcomment}

\usepackage{multirow}
\usepackage{setspace}
\usepackage{color}

\usepackage{float}
\usepackage[T1]{fontenc}
\usepackage[latin9]{inputenc}
\usepackage{setspace}
\usepackage{esint}

\providecommand{\tabularnewline}{\\}

\begin{document}

%\preprint{APS/123-QED}

\title{Entanglement phase transitions in non-Hermitian quasicrystals}

\author{Longwen Zhou}
\email{zhoulw13@u.nus.edu}
\affiliation{%
	College of Physics and Optoelectronic Engineering, Ocean University of China, Qingdao, China 266100
}
\affiliation{%
	Key Laboratory of Optics and Optoelectronics, Qingdao, China 266100
}
\affiliation{%
	Engineering Research Center of Advanced Marine Physical Instruments and Equipment of MOE, Qingdao, China 266100
}

\date{\today}

\begin{abstract}
The scaling law of entanglement entropy could undergo qualitative
changes during the nonunitary evolution of a quantum many-body system.
In this work, we uncover such entanglement phase transitions for free fermions in one-dimensional
non-Hermitian quasicrystals (NHQCs). We identify two types of entanglement
transitions with different scaling laws and critical behaviors due
to the interplay between non-Hermitian effects and quasiperiodic potentials.
The first type represents a typical volume-law to area-law transition,
which happens together with a PT-symmetry breaking and a localization
transition. The second type features an abnormal log-law to area-law
transition, which is mediated by a critical phase with a volume-law
scaling in the steady-state entanglement entropy. These entangling
phases and transitions are demonstrated in two representative models
of noninteracting NHQCs. Our results thus advanced the study of entanglement transitions
in non-Hermitian disordered systems and further disclosed the rich
entanglement patterns in NHQCs.
\end{abstract}

%Use showkeys class option if keyword
\maketitle

\section{Introduction\label{sec:Int}}

Along with the increase of measurement rates, the competition between
unitary time evolution and projective measurements could prompt the
steady state of a quantum many-body system (either interacting or noninteracting) to switch from a volume-law-entangled
phase to a quantum Zeno phase with an area-law entanglement entropy
(EE) \cite{EPT01,EPT02,EPT03,EPT04,EPT05}. Ever since its discovery,
this measurement-induced entanglement phase transition has attracted
great attention in both theoretical \cite{EPT06,EPT07,EPT08,EPT09,EPT10,EPT11,EPT12,EPT13,EPT14,EPT15,EPT16,EPT17,EPT18,EPT19,EPT20,EPT21,EPT22,EPT23,EPT24,EPT25,EPT26,EPT27,EPT28,EPT29,EPT30,EPT31,EPT32,EPT33,EPT34,EPT35,EPT36,EPT37,EPT38,EPT39,EPT40,EPT41,EPT42,EPT43,EPT44,EPT45,EPT46,EPT47,EPT48,EPT49,EPT50,EPT51,EPT52}
and experimental \cite{EPTExp1,EPTExp2,EPTExp3} studies, with important implications
for the understanding of quantum information dynamics and the simulation
of quantum many-body systems \cite{EPTRev1,EPTRev2,EPTRev3}. Recently,
entanglement phase transitions are also studied in the context
of non-Hermitian physics \cite{NHEPT01,NHEPT02,NHEPT03,NHEPT04,NHEPT05,NHEPT06,NHEPT07,NHEPT08}.
There, the effect of measurement is taken into account by considering
a nonunitary evolution generated by a non-Hermitian Hamiltonian. The
dissipation-gap formation and the non-Hermitian skin effect are further
identified as two typical mechanisms of producing entangling-disentangling
phase transitions \cite{NHEPT04,NHEPT05}. Yet, these discoveries
are established with a focus on pristine non-Hermitian lattice models.

Non-Hermitian quasicrystal (NHQC) forms a typical category of disordered
non-Hermitian setup \cite{NHQC01,NHQC02,NHQC03}. In an NHQC, the
interplay between correlated disorder and gain/loss or nonreciprocal
effects could yield rich phases and phenomena including parity-time-reversal- (PT-) symmetry
breaking transitions, localization transitions, topological transitions
and mobility edges \cite{NHQC04,NHQC05,NHQC06,NHQC07,NHQC08,NHQC09}.
Despite great theoretical efforts \cite{NHQC10,NHQC11,NHQC12,NHQC13,NHQC14,NHQC15,NHQC16,NHQC17,NHQC18,NHQC19,NHQC20,NHQC21,NHQC22,NHQC23,NHQC24,NHQC25,NHQC26,NHQC27,NHQC28,NHQC29,NHQC30,NHQC31,NHQC32,NHQC33,NHQC34,NHQC35,NHQC36,NHQC37,NHQC38,NHQC39,NHQC40,NHQC41},
NHQCs have also been experimentally realized by nonunitary photonic
quantum walks \cite{NHQCExp1,NHQCExp2}. However, much less is known
regarding entanglement phase transitions in NHQCs \cite{NHQC38,NHQC39}.
This question can be interesting, as a PT-broken NHQC could belong
to either a localized phase \cite{NHQC05} or an extended phase \cite{NHQC04}.
In the latter case, the delocalized bulk states should prefer a volume-law
scaling in the steady-state EE after a long-time evolution, while
the dissipation gap in the complex energy spectrum may favor an area-law
entanglement scaling. The competition between these two opposite trends
may lead to new scaling laws and exotic critical behaviors for the
EE. Moreover, an NHQC could possess a point-gap instead of a line-gap
on the complex energy plane \cite{NHQC04,NHQC05}, and the implication
of a point dissipation gap on entanglement phase transitions is largely
unclear. Besides, whether and how entanglement transitions would accompany
other phase transitions (e.g., PT-symmetry breaking, localization,
etc.) in NHQCs remain to be uncovered.

To resolve these puzzles, we explore in this work the entanglement
phase transitions of free fermions in NHQCs, with a focus on two ``minimal'' and
mutually dual non-Hermitian lattice models \cite{NHQC06,NHQC13}.
In Sec.~\ref{sec:Mod}, we introduce these models and review their
known spectral and localization properties. The entanglement phase
transitions in these models are explored in detail in Sec.~\ref{sec:Res}.
A unique type of log-law to area-law entanglement transition, mediated
by a volume-law critical entangling phase is identified. In Sec.~\ref{sec:Sum},
we summarize our results, comment on related issues and discuss potential future directions.

\section{Models\label{sec:Mod}}

We focus on the entanglement phase transitions in two ``minimal''
non-Hermitian variants of the noninteracting Aubry-Andr\'e-Harper (AAH) model. They
will be denoted by NHAAH1 and NHAAH2 for brevity. We first go over
some of their key physical properties in this section. Throughout
the discussions, we will set the lattice constant $a=1$ and the Planck
constant $\hbar=1$.

In the position representation, the Hamiltonian of the NHAAH1 takes the
form of
\begin{equation}
	\hat{H}_{1}=J\sum_{n}(\hat{c}_{n}^{\dagger}\hat{c}_{n+1}+\hat{c}_{n+1}^{\dagger}\hat{c}_{n})+V\sum_{n}e^{-i2\pi\alpha n}\hat{c}_{n}^{\dagger}\hat{c}_{n}.\label{eq:H1}
\end{equation}
Here $\hat{c}_{n}^{\dagger}$ ($\hat{c}_{n}$) creates (annihilates)
a fermion on the lattice site $n$. $J$ denotes the nearest-neighbor
hopping amplitude and $V$ denotes the amplitude of onsite potential
$V_{n}=Ve^{-i2\pi\alpha n}$. 
$2\pi\alpha$ describes the wavenumber of the superlattice \cite{Note0}.
Expanding a general state as $|\psi\rangle=\sum_{n}\psi_{n}\hat{c}_{n}^{\dagger}|\emptyset\rangle$,
the eigenvalue equation $\hat{H}_{1}|\psi\rangle=E|\psi\rangle$ of NHAAH1 can
be expressed in the following form
\begin{equation}
	J\psi_{n+1}+J\psi_{n-1}+Ve^{-i2\pi\alpha n}\psi_{n}=E\psi_{n}.\label{eq:Seq1}
\end{equation}
Here $|\emptyset\rangle$ denotes the vacuum state and the amplitude
$\psi_{n}$ is normalized as $\sum_{n}|\psi_{n}|^{2}=1$. It is clear
that the NHAAH1 is non-Hermitian due to the complex onsite phase factor
$e^{-i2\pi\alpha n}$. It further possesses the PT symmetry, with
the parity ${\cal P}:n\rightarrow-n$ and the time-reversal ${\cal T}={\cal K}$,
where ${\cal K}$ performs the complex conjugation. The quasicrystal
nature of the system comes about by setting $\alpha$ as an irrational
number, so that the onsite potential is spatially quasiperiodic. The
energy spectrum of the system under the periodic boundary condition
(PBC) was found to take the conjectured form of \cite{NHQC06}
\begin{equation}
	E=\begin{cases}
		2J\cos k & |V|\leq|J|\\
		\left(V+\frac{J^{2}}{V}\right)\cos k+i\left(V-\frac{J^{2}}{V}\right)\sin k & |V|>|J|
	\end{cases}.\label{eq:H1E}
\end{equation}
Here $k\in[-\pi,\pi)$ is an artificial parameter that tells us the eigenenergies either fill the region of a line segment or an ellipse \cite{Note0x}. Therefore, the spectrum is real for $|V|<|J|$
(PT-invariant) and complex for $|V|>|J|$ (PT-broken). There is a PT transition in the energy spectrum at $|V|=|J|$.

The Hamiltonian of the NHAAH2 in the position representation is given
by
\begin{equation}
	\hat{H}_{2}=J\sum_{n}\hat{c}_{n+1}^{\dagger}\hat{c}_{n}+2V\sum_{n}\cos(2\pi\alpha n)\hat{c}_{n}^{\dagger}\hat{c}_{n},\label{eq:H2}
\end{equation}
and the related eigenvalue equation reads
\begin{equation}
	J\psi_{n-1}+2V\cos(2\pi\alpha n)\psi_{n}=E\psi_{n}.\label{eq:Seq2}
\end{equation}
It is clear that the nearest-neighbor hopping is unidirectional from
left to right, making the system non-Hermitian. 
The NHAAH1 and NHAAH2 are differ in both their hopping and onsite potential terms, so that neither of them can be viewed as a special case of the other.
The NHAAH2
is also quasiperiodic if $\alpha$ is irrational \cite{Note0}. Taking
a rational approximation for $\alpha\simeq p/q$ (with $p,q$ being
co-prime integers) and performing the discrete Fourier transformation
$\psi_{n}=\frac{1}{L}\sum_{\ell=1}^{L}\phi_{\ell}e^{i2\pi\alpha\ell n}$
under the PBC ($\psi_{n}=\psi_{n+L}$), the Eq.~(\ref{eq:Seq2}) can
be transformed to the momentum space \cite{NHQC05} as
\begin{equation}
	V\phi_{\ell+1}+V\phi_{\ell-1}+Je^{-i2\pi\alpha\ell}\phi_{\ell}=E\phi_{\ell},\label{eq:Seq3}
\end{equation}
where $L$ denotes the length of lattice. It is now clear that the
NHAAH2 also possesses the PT symmetry, with ${\cal P}:\ell\rightarrow-\ell$
and ${\cal T}={\cal K}$. The energy spectrum of the system under
the PBC is further given by the conjectured expression
\begin{equation}
	E=\begin{cases}
		\left(J+\frac{V^{2}}{J}\right)\cos k+i\left(J-\frac{V^{2}}{J}\right)\sin k & |V|<|J|\\
		2V\cos k & |V|\geq|J|
	\end{cases},\label{eq:H2E}
\end{equation}
where $k\in[-\pi,\pi)$ is again an artificial parameter \cite{Note0x}. Therefore, the spectrum is real for $|V|>|J|$
(PT-invariant) and complex for $|V|<|J|$ (PT-broken). The PT transition
of NHAAH2 also happens at $|V|=|J|$ \cite{NHQC13}.

By comparing the Eqs.~(\ref{eq:Seq1}) and (\ref{eq:Seq3}), we further
observe a duality relation between the NHAAH1 and NHAAH2, implying
the presence of a fixed point along $|J|=|V|$. In fact, it has been
identified that under the PBC and for any irrational $\alpha$, there
is a PT spectral transition together with a localization-delocalization
transition at $|J|=|V|$ for both the NHAAH1 and NHAAH2. When $|V|<|J|$,
the NHAAH1 (NHAAH2) resides in an extended phase with a real (complex)
spectrum and holding only extended eigenstates. When $|V|>|J|$, the
NHAAH1 (NHAAH2) switches to a localized phase with a complex (real)
spectrum and holding only localized eigenstates \cite{NHQC06,NHQC13}. The transitions between
these phases could be further captured by quantized changes of spectral
topological winding numbers \cite{NHQC04,NHQC05}.

\begin{figure}
	\begin{centering}
		\includegraphics[scale=0.475]{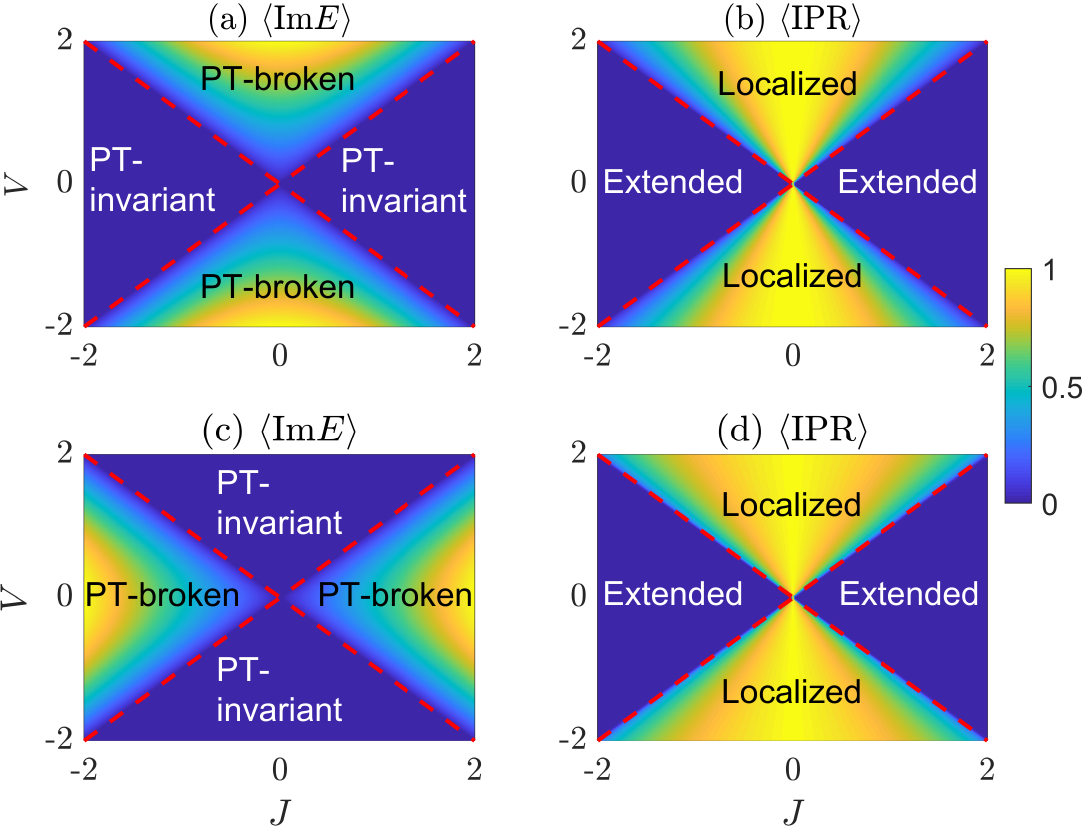}
		\par\end{centering}
	\caption{Phase diagrams of the NHAAH1 {[}in (a), (b){]} and NHAAH2 {[}in (c),
		(d){]} under the PBC \cite{Note1}. We choose $\alpha=\frac{\sqrt{5}-1}{2}$
		and the length of lattice $L=610$ for all panels. The red dashed
		lines show the phase boundaries $J=\pm V$. The NHAAH1 stays in a
		PT-invariant extended phase for $|V|<|J|$ and goes to a PT-broken
		localized phase for $|V|>|J|$. The NHAAH2 resides in a PT-broken
		extended phase for $|V|<|J|$ and switches to a PT-invariant localized
		phase for $|V|>|J|$. \label{fig:E-IPR}}
\end{figure}
In Fig.~\ref{fig:E-IPR}, we illustrate the phases and transitions
in NHAAH1 and NHAAH2 by investigating their spectra and inverse
participation ratios (IPRs). The $\langle{\rm Im}E\rangle$ {[}in
Figs.~\ref{fig:E-IPR}(a) and \ref{fig:E-IPR}(c){]} and $\langle{\rm IPR}\rangle$
{[}in Figs.~\ref{fig:E-IPR}(b) and \ref{fig:E-IPR}(d){]} are defined
as
\begin{equation}
	\langle{\rm Im}E\rangle=\frac{1}{L}\sum_{j=1}^{L}|{\rm Im}E_{j}|,\label{eq:ImE}
\end{equation}
\begin{equation}
	\langle{\rm IPR}\rangle=\frac{1}{L}\sum_{j=1}^{L}\sum_{n=1}^{L}|\psi_{n}^{j}|^{4}.\label{eq:IPR}
\end{equation}
Here $E_{j}$ is the $j$th eigenenergy of $\hat{H}_{1}$ or $\hat{H}_{2}$
with the normalized right eigenvector $|\psi_{j}\rangle=\sum_{n=1}^{L}\psi_{n}^{j}\hat{c}_{n}^{\dagger}|\emptyset\rangle$.
By definition, we expect $\langle{\rm Im}E\rangle=0$ ($\langle{\rm Im}E\rangle>0$)
in the PT-invariant (PT-broken) phase, and $\langle{\rm IPR}\rangle\rightarrow0$
($\langle{\rm IPR}\rangle>0$) in the extended (localized) phase.
The numerical results presented in Fig.~\ref{fig:E-IPR} clearly verified
the theoretically predicted extended/localized phases, PT transitions
and localization transitions in these NHQCs \cite{NHQC06,NHQC13}.

Based on these known physical properties, one may expect to have entanglement
phase transitions also in the NHAAH1 and NHAAH2. For example, after a long-time
evolution, the EE of a typical initial state might be proportional
to the system size (volume-law) in the PT-invariant phase and become
almost independent to the system size in the PT-broken phase (area-law)
\cite{NHEPT05}. The PT transition of NHAAH1 or NHAAH2 should then
accompany a volume-law to area-law entanglement transition. Meanwhile,
one may also expect the scaling of steady-state EE to follow a volume-law
in the extended phase and an area-law in the localized phase. However,
the PT-invariant (PT-broken) phase of our system could also be a localized
(an extended) phase, making the real physical situation more complicated.
As will be demonstrated in the following section, despite the more
conventional volume-law to area-law entanglement transitions, the
steady-state EE of an NHQC may follow an abnormal log-law scaling due to the interplay
between quasiperiodicity and non-Hermitian effects. A log-law to area-law
entanglement phase transition could further be induced to happen across
a critical point where the EE follows a volume-law.

\section{Results\label{sec:Res}}
In this section, we reveal the entanglement phase transitions in NHAAH1
and NHAAH2. We first discuss the definition of EE and the calculation
of its dynamics for a non-Hermitian system. Next, we demonstrate the
scaling relations of steady-state EE with respect to the system and
subsystem sizes for our two NHQC models in Secs.~\ref{subsec:NHAAH1}
and \ref{subsec:NHAAH2}. These relations allow us to clearly distinguish
different entangling phases in the considered systems. The entanglement
phase transitions are further uncovered by investigating the changes
of EE with respect to different system parameters. With all these
information, we finally establish the entanglement phase diagrams
for our NHQC models.

For a system consists of noninteracting fermions, the EE of an arbitrary
subsystem and its time evolution can be extracted from the spectrum and
dynamics of the single-particle correlator. Consider a system described
by the quadratic Hamiltonian $\hat{H}=\sum_{m,n}\hat{c}_{m}^{\dagger}H_{mn}\hat{c}_{n}$
and prepared at time $t=0$ in the initial state $|\Psi_{0}\rangle$,
the normalized state of the system at a later time $t$ is given by
\begin{equation}
	|\Psi(t)\rangle=\frac{e^{-i\hat{H}t}|\Psi_{0}\rangle}{\sqrt{\langle\Psi_{0}|e^{i\hat{H}^{\dagger}t}e^{-i\hat{H}t}|\Psi_{0}\rangle}}.\label{eq:Psit}
\end{equation}
Here $\hat{c}_{m}^{\dagger}$ ($\hat{c}_{n}$) creates (annihilates)
a fermion at the lattice site $m$ ($n$). Note that for a non-Hermitian
system, we generally have $\hat{H}\neq\hat{H}^{\dagger}$, leading
to a nonunitary time evolution. In our calculations, we choose the
initial state to be in the form of a charge density wave for a half-filled
lattice, i.e.,
\begin{equation}
	|\Psi_{0}\rangle=\prod_{r\in\mathbb{Z}}\hat{c}_{2r}^{\dagger}|\emptyset\rangle,\label{eq:Psi0}
\end{equation}
where $r=1,2,...,\lfloor L/2\rfloor-1,\lfloor L/2\rfloor$. Other
kinds of pure initial states generate similar results regarding the (sub)system-size
scaling of steady-state EE. At a later time $t$, the element of single-particle
correlation matrix $C(t)$ in position representation is given by
\begin{equation}
	C_{mn}(t)=\langle\Psi(t)|\hat{c}_{m}^{\dagger}\hat{c}_{n}|\Psi(t)\rangle,\label{eq:CMt}
\end{equation}
Restricting the indices $m$ and $n$ to a subsystem A of size $l$
and diagonalizing the corresponding $l\times l$ block of $C(t)$
result in the correlation-matrix spectrum $\{\zeta_{j}(t)|j=1,...,l\}$. The EE at time
$t$ can then be obtained as \cite{NHEPT04}
\begin{equation}
	S(t)=-\sum_{j=1}^{l}[\zeta_{j}(t)\ln\zeta_{j}(t)+(1-\zeta_{j}(t))\ln(1-\zeta_{j}(t))].\label{eq:St}
\end{equation}
Note that the $S(t)$ here is the bipartite EE of a subsystem A. It is
defined by tracing over all the degrees of freedom belonging to a
complementary subsystem B of the size $L-l$, in the sense that $S=-{\rm Tr}(\rho_{{\rm A}}\ln\rho_{{\rm A}})$
and $\rho_{{\rm A}}={\rm {\rm Tr}_{B}}(|\Psi(t)\rangle\langle\Psi(t)|)$.
Numerically, the EE of a Gaussian state can be computed efficiently
following the recipe listed in the Appendix B of Ref.~\cite{NHEPT04}.

In the following subsections, we study the EE of our two NHQC models
with the method outlined here. We focus on systems under the PBC and
set the irrational parameter $\alpha=(\sqrt{5}-1)/2$ (the inverse
golden ratio) for all our calculations. Other choices of the irrational
$\alpha$ lead to similar results.

\begin{figure*}
	\begin{centering}
		\includegraphics[scale=0.35]{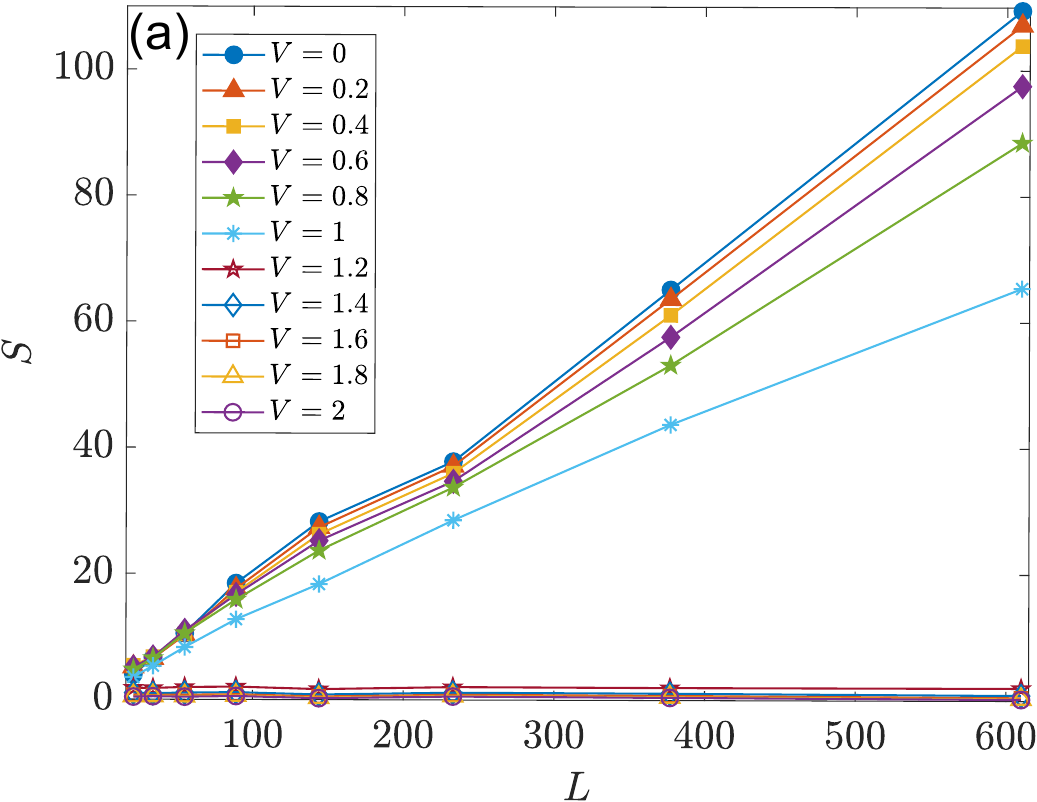}$\,\,$\includegraphics[scale=0.35]{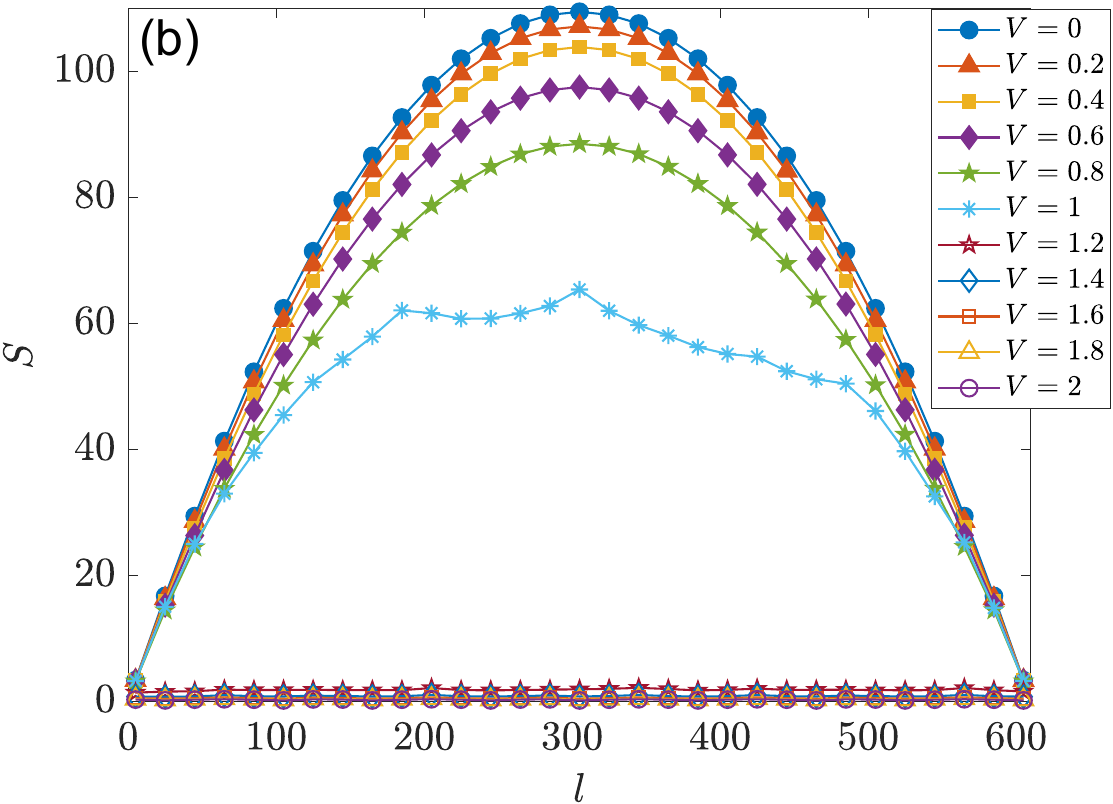}
		\par\end{centering}
	\begin{centering}
		\includegraphics[scale=0.35]{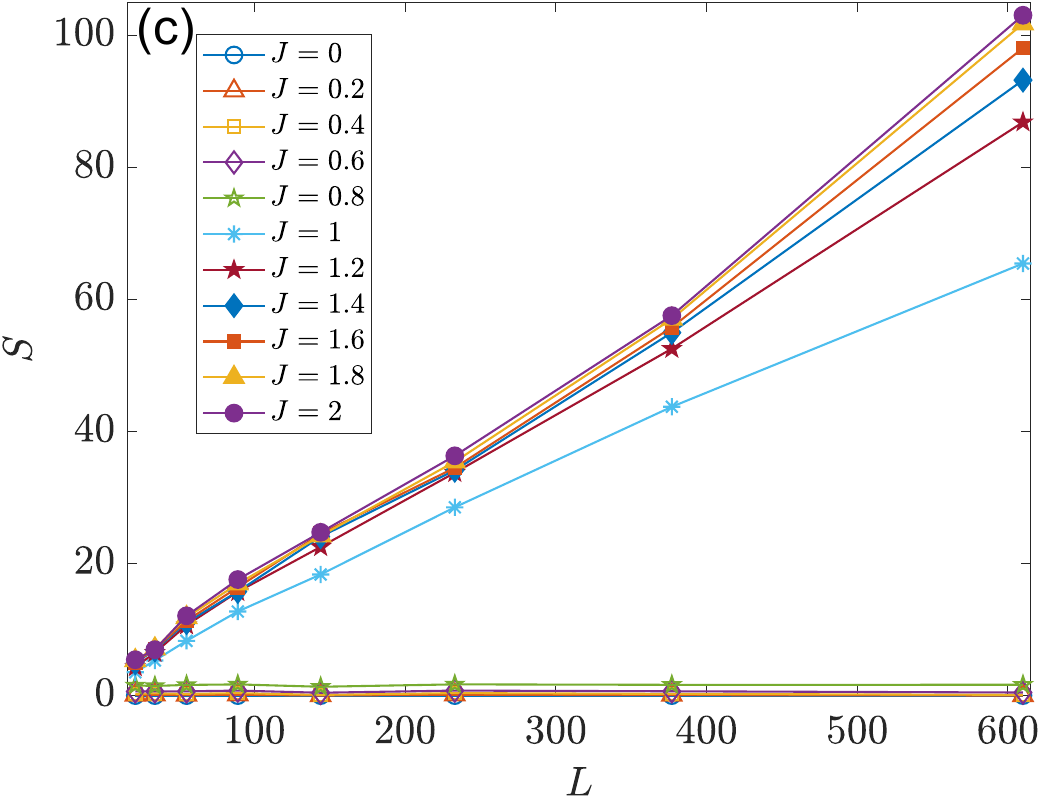}$\,\,$\includegraphics[scale=0.35]{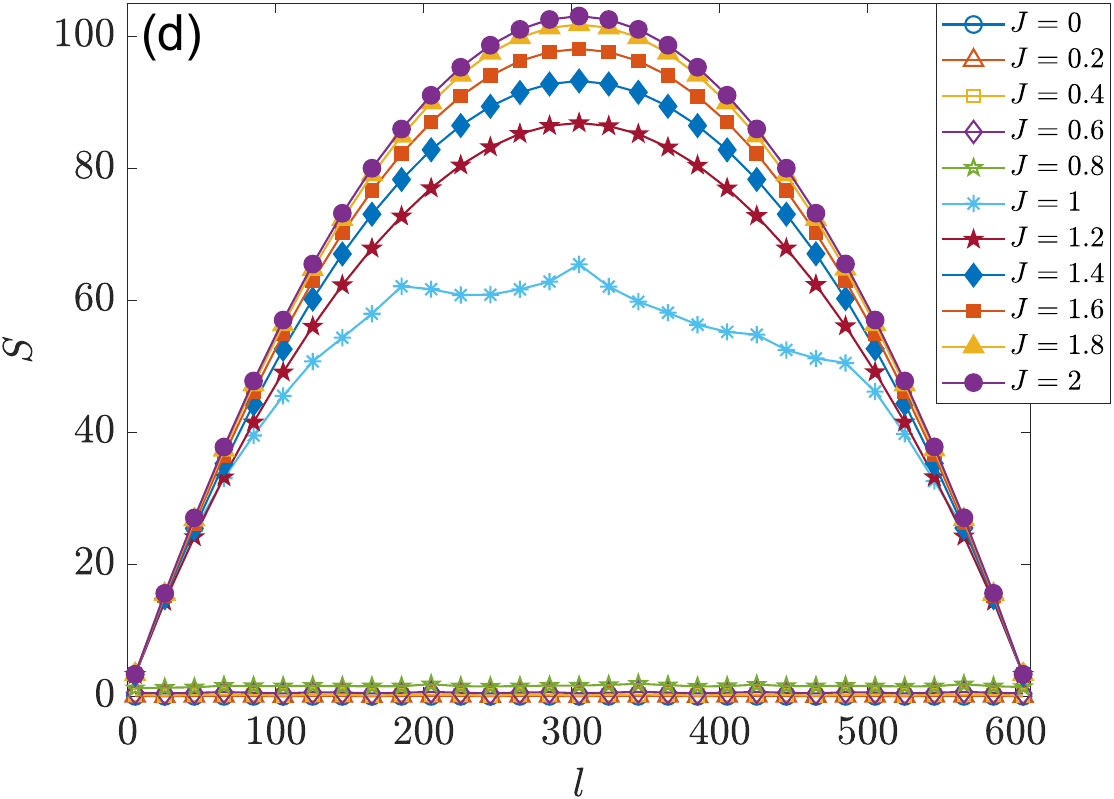}
		\par\end{centering}
	\caption{Steady-state EE of the NHAAH1 at half-filling versus the system size
		$L$ {[}under bi-partition in (a), (c){]} and the subsystem size $l$
		{[}under a fixed length of lattice $L=610$ in (b), (d){]}. Other
		system parameters are set as $J=1$ for (a), (b) and $V=1$ for (c),
		(d). The time span of the entire evolution is $T=1000$ \cite{Note2}. \label{fig:EEvsLLS1}}
\end{figure*}

\subsection{NHAAH1\label{subsec:NHAAH1}}

We first reveal entanglement phase transitions in the NHAAH1 by investigating
its steady-state EE $S(L,l)$, with $L$ and $l$ being the length of lattice
and the size of its subsystem A. The system is prepared at
$t=0$ in the initial state $|\Psi_{0}\rangle$ {[}Eq.~(\ref{eq:Psi0}){]}
and then evolved according to Eq.~(\ref{eq:Psit}), with the $\hat{H}$
given by Eq.~(\ref{eq:H1}). The EE $S(t)$ at a later time $t$ 
{[}Eq.~(\ref{eq:St}){]} is obtained from the spectrum of correlation matrix
$C(t)$ {[}Eq.~(\ref{eq:CMt}){]} restricted to the subsystem A. Focusing
on a long-time evolution of duration $T$, we obtain the steady-state
EE $S(L,l)$ by averaging $S(t)$ over a suitable time window $t\in[T',T]$
with $1\ll T'<T$. The scaling property of $S(L,l)$ can then be analyzed
by considering different choices of $L$ and $l$.

\begin{figure*}
	\begin{centering}
		\includegraphics[scale=0.35]{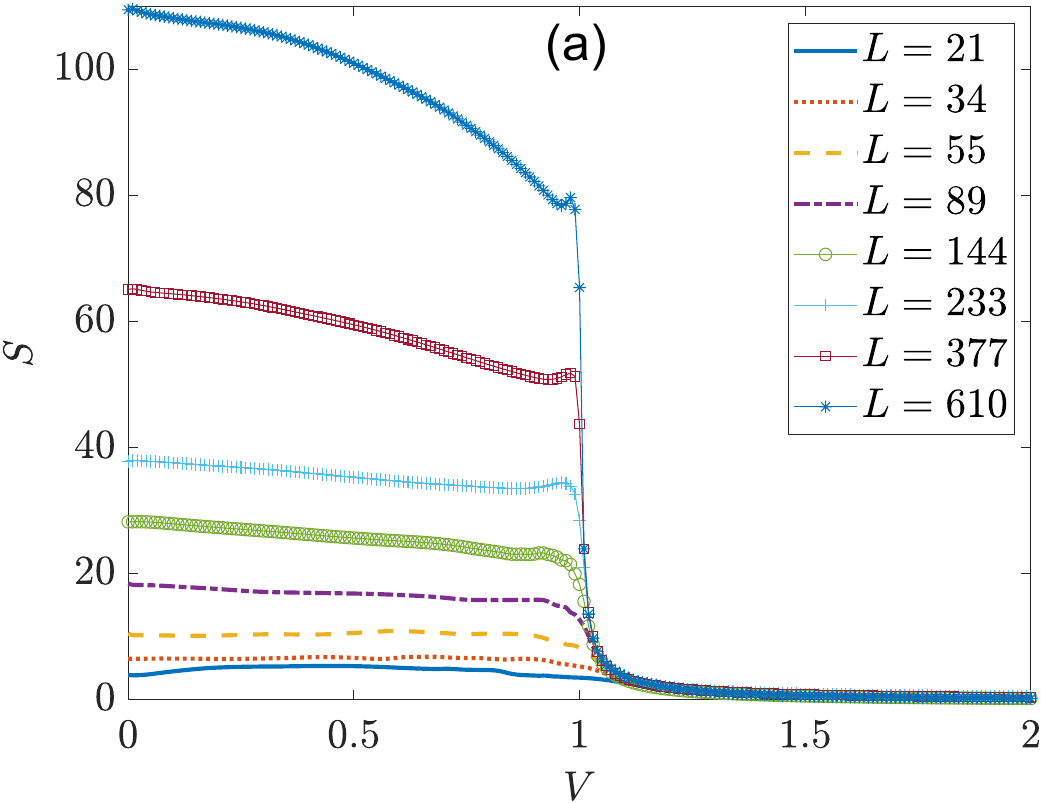}$\,\,$\includegraphics[scale=0.35]{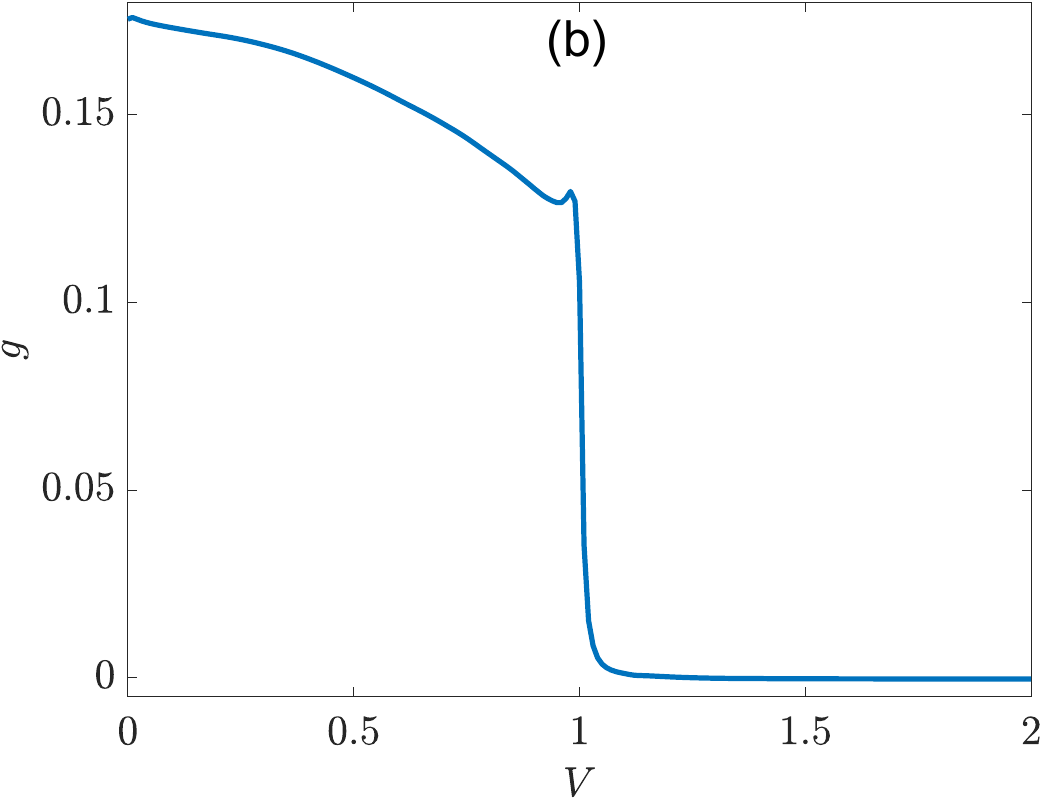}
		\par\end{centering}
	\begin{centering}
		\includegraphics[scale=0.35]{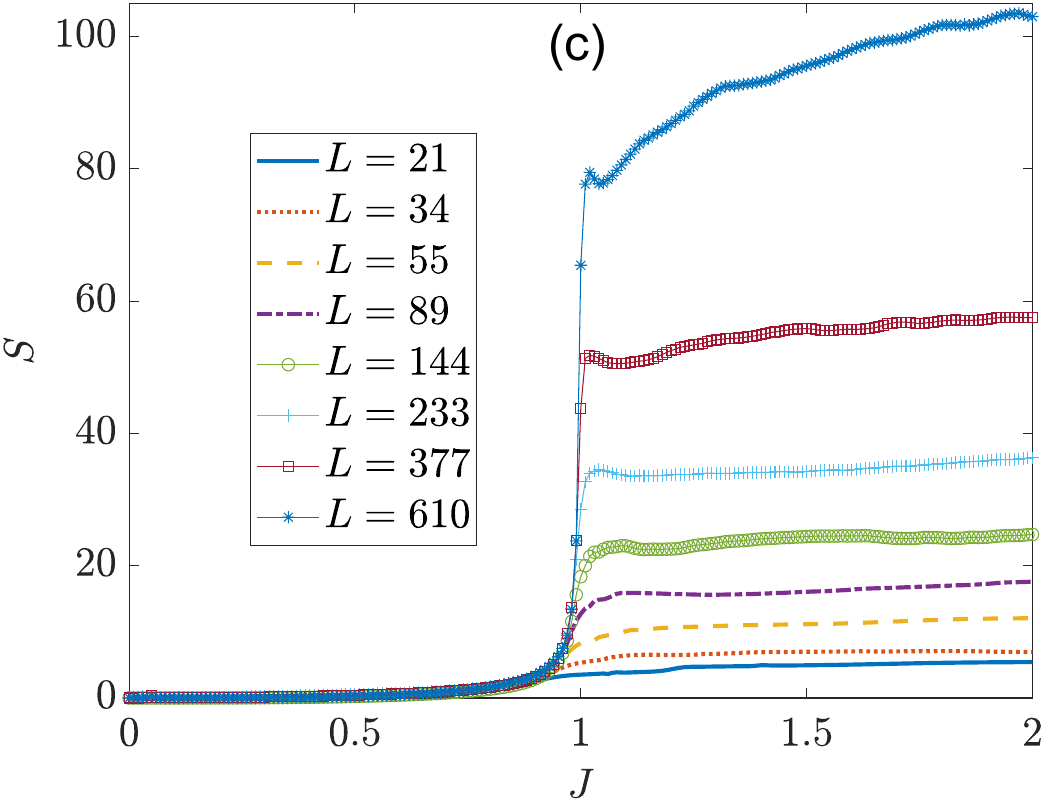}$\,\,$\includegraphics[scale=0.35]{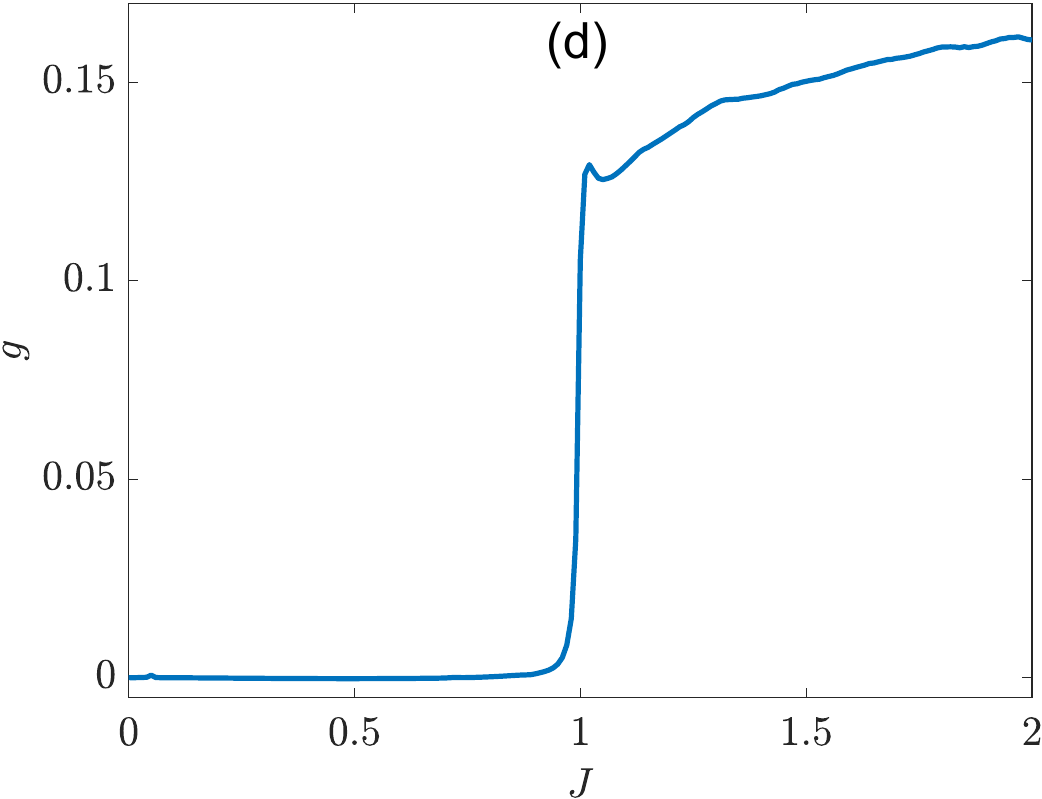}
		\par\end{centering}
	\caption{Bipartite EE of the steady state at half-filling {[}in (a), (c){]}
		and the related gradient $g$ in the scaling law of steady-state EE {[}in (b) and
		(d){]} for the NHAAH1. Other system parameters are set as $J=1$ for
		(a), (b) and $V=1$ for (c), (d). The time span of the entire evolution
		is $T=1000$ \cite{Note2}. The values of $g$ are obtained from the linear fitting
		$S(L,L/2)\sim gL+s_{0}$ of EE versus the lattice size $L$ at given system
		parameters. \label{fig:EEvsJV1}}
\end{figure*}

In Fig.~\ref{fig:EEvsLLS1}, we present the steady-state EE versus
the system size $L$ and the subsystem size $l$ for typical sets
of system parameters. In Figs.~\ref{fig:EEvsLLS1}(a) and \ref{fig:EEvsLLS1}(c), we
consider a equal bipartition of the system ($l=\lfloor L/2\rfloor$). For $|V|>|J|$,
we find that the $S(L,L/2)$ almost does not change with $L$, which
implies that the PT-broken localized phase of the NHAAH1 is area-law
entangled. This is expected, as in this case the point dissipation
gap on the complex energy plane {[}see Eq.~(\ref{eq:H1E}){]} and
the spatial localization of all eigenstates both tend to hinder the
spreading of quantum entanglement across the system. For $|V|<|J|$,
we instead observe that up to leading order, the EE is proportional
to the system size $L$, i.e., $S(L,L/2)\propto gL$ with the gradient
$g\approx0.1\sim0.2$. Therefore, in the PT-invariant extended phase
of NHAAH1, the steady-state EE tends to satisfy a volume-law. Such
a linear scaling is triggered by the quantum information spreading
due to delocalized bulk states with real energies in the system. The
gradient $g$ of the volume-law scaling decreases gradually but remains
finite till the critical point of PT and localization transitions
at $|J|=|V|$. 

In Figs.~\ref{fig:EEvsLLS1}(b) and \ref{fig:EEvsLLS1}(d), we consider a fixed system
size $L$ and obtain the curve $S(L,l)$ vs the size $l$ of subsystem
A for $l\in(0,L)$. The results show that for $|V|>|J|$, the $S(L,l)$
is almost independent of $l$ up to slight fluctuations, which is
an expected situation for an area-law entangled phase. For $|V|<|J|$,
the $S(L,l)$ as a function of $l$ can be numerically fitted as
$S(L,l)\simeq A\sin(\pi l/L)+B\ln[\sin(\pi l/L)]+C$, where
$A$, $B$ and $C$ are some fitting coefficients. This is typical
for a volume-law entangled phase. Putting together, we conclude that
the steady-state EE of NHAAH1 indeed follows qualitatively different
scaling laws with respect to the (sub)system size in different parameter
regions, which implies the presence of entanglement phase transitions
in the system.

To further decipher the entanglement transitions in NHAAH1, we present
its steady-state EE $S(L,L/2)$ versus $V$ and $J$ for different
system sizes $L$ in Figs.~\ref{fig:EEvsJV1}(a) and \ref{fig:EEvsJV1}(c). Two distinct
regions can be clearly figured out. In the region with $|J|<|V|$,
the EE shows an $L$-independence. Whereas for $|J|>|V|$, the EE
increases monotonically with $L$. A marked change is observed at
$|J|=|V|$ in the $L$-dependence of $S(L,L/2)$, which implies a transition
in the scaling law of EE. In Figs.~\ref{fig:EEvsJV1}(b) and \ref{fig:EEvsJV1}(d),
we obtain the gradient $g$ by fitting the steady-state EE $S(L,L/2)$
with the function $gL+s_{0}$ at different values of $J$ and $V$.
The results show that $g\simeq0$ {[}$S(L,L/2)\sim s_{0}\sim L^0${]}
for $|J|<|V|$ and $g>0$ {[}$S(L,L/2)\propto L${]} for $|J|>|V|$,
which are expected behaviors for area-law entangled and volume-law
entangled phases, respectively. There is then a discontinuous change
of $g$ at $|J|=|V|$, which signifies an entanglement phase transition
in the NHAAH1.

\begin{figure}
	\begin{centering}
		\includegraphics[scale=0.47]{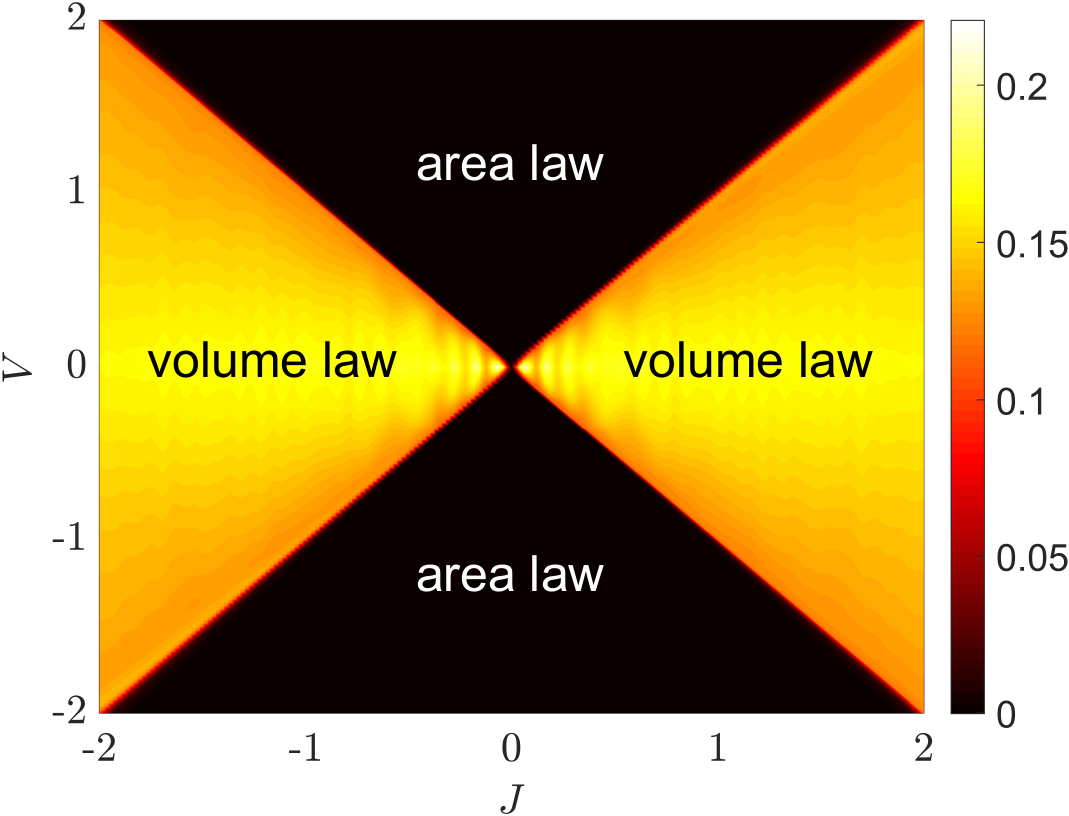}
		\par\end{centering}
	\caption{Entanglement phase diagram of the NHAAH1. Different colors correspond
		to different values of the gradient $g$ extracted from the linear
		fitting $S(L,L/2)\sim gL+s_{0}$ of steady-state EE versus the system size
		$L$. \label{fig:EEPhsDiag1}}
\end{figure}

The physics behind the different scaling laws of EE is as follows. For our NHAAH1, initial excitations could propagate and spread uniformly across the whole lattice with the increase of time for $|V|<|J|$, i.e., in the PT-invariant extended phase, and the entanglement is building up throughout the system before reaching a steady state. When the steady-state is reached, an extensive entanglement is retained across any spatial cuts in the lattice. The resulting bipartite EE then follows a volume-law ($gL$) vs the system size. For $|V|>|J|$, all the eigenstates are localized, and any initial excitations could not propagate and spread in the system after an initial transient time window. Moreover, the onsite gain and loss are strong enough when $|V|>|J|$ so as to disentangle degrees of freedom at different spatial locations. Therefore, any extensive entanglement could not be established across the system due to the collaboration between strong disorder and strong gain/loss (which may also be viewed as strong measurement backactions \cite{NHEPT05}). The result is an area-law scaling of steady-state EE versus the system size in the PT-broken localized phase of our system.

\begin{table*}
	\begin{centering}
		\begin{tabular}{|c|c|c|c|}
			\hline 
			NHAAH1 & $|V|<|J|$ & $|V|=|J|$ & $|V|>|J|$\tabularnewline
			\hline 
			\hline 
			Energy spectrum & real & PT transition & complex\tabularnewline
			\hline 
			Eigenstates & extended & localization transition & localized\tabularnewline
			\hline 
			Steady-state EE & volume-law & entanglement transition & area-law\tabularnewline
			\hline 
		\end{tabular}
		\par\end{centering}
	\caption{Summary of main results for the quasicrystal NHAAH1 \ref{eq:H1}. The real-spectrum
		(PT-invariant), extended phase is volume-law entangled. The complex-spectrum
		(PT-broken), localized phase is area-law entangled. The PT, localization
		and entanglement phase transitions happen all together at $|V|=|J|$ [see also Figs.~\ref{fig:E-IPR}(a), \ref{fig:E-IPR}(b) and \ref{fig:EEPhsDiag1}].}\label{Tab1}
\end{table*}

Collecting together the scaling properties of steady-state EE with
respect to the lattice size $L$ for a half-filled and bipartite system,
we arrive at the entanglement phase diagram of NHAAH1 under the PBC
in Fig.~\ref{fig:EEPhsDiag1}. 
A summary of the key features of NHAAH1 is given in Table \ref{Tab1}.
We find that there are indeed two phases
with different entanglement nature, which are separated by an entanglement
transition at $|J|=|V|$. In the PT-broken localized phase ($|J|<|V|$),
the system is found to be area-law entangled {[}$S(L,L/2)\sim L^0${]}.
The spectrum is complex with a point dissipation gap at $E=0$ on
the complex energy plane {[}see Eq.~(\ref{eq:H1E}){]} and all the
eigenstates are localized, both compelling the termination of entanglement
spreading in this case. In the PT-invariant extended phase ($|J|>|V|$),
the system is instead volume-law entangled {[}$S(L,L/2)\propto L$
up to the leading order{]}. Since the system possesses a real spectrum
{[}Eq.~(\ref{eq:H1E}){]} and all its eigenstates are extended in
this case, the quantum information is forced to spread and a volume-law
entangled phase results. Such a volume-law to area-law entanglement
phase transition was identified before in clean non-Hermitian systems
due to different physical mechanisms \cite{NHEPT04,NHEPT05}. 
In the next subsection, we will demonstrate that an even more exotic type
of entanglement phase transition could emerge in NHQCs due to the
interplay between disorder and nonreciprocity.

\begin{figure*}
	\begin{centering}
		\includegraphics[scale=0.35]{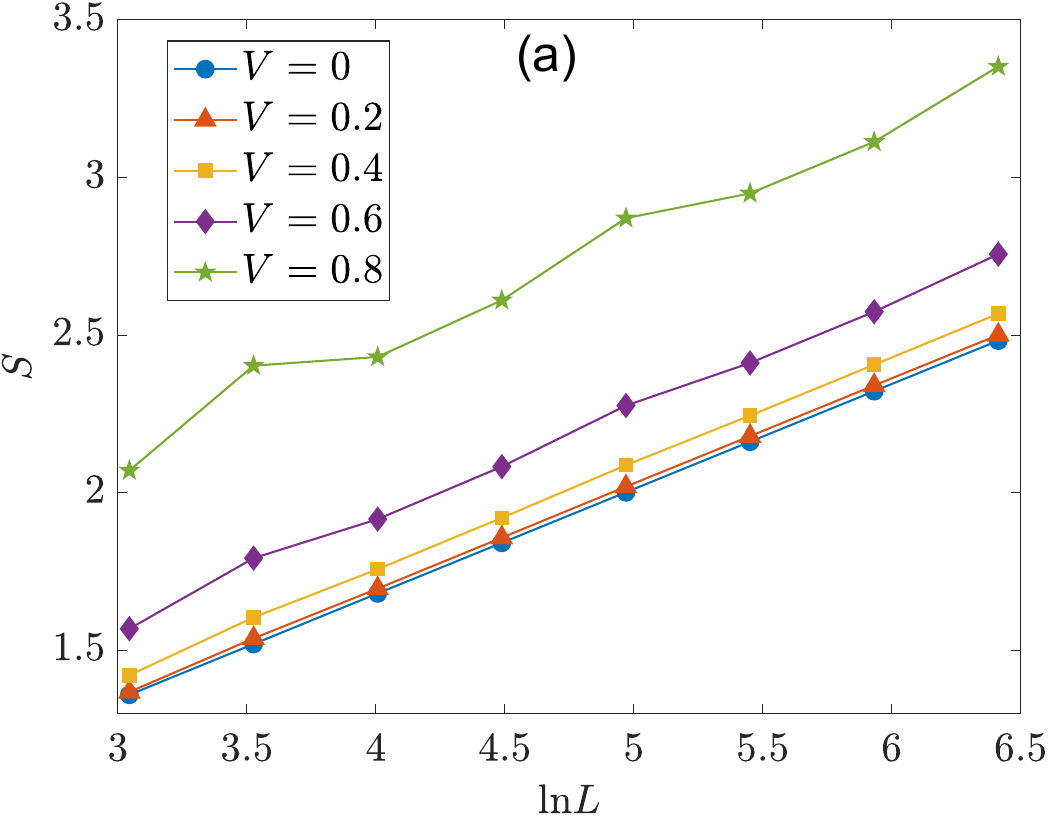}$\,$\includegraphics[scale=0.226]{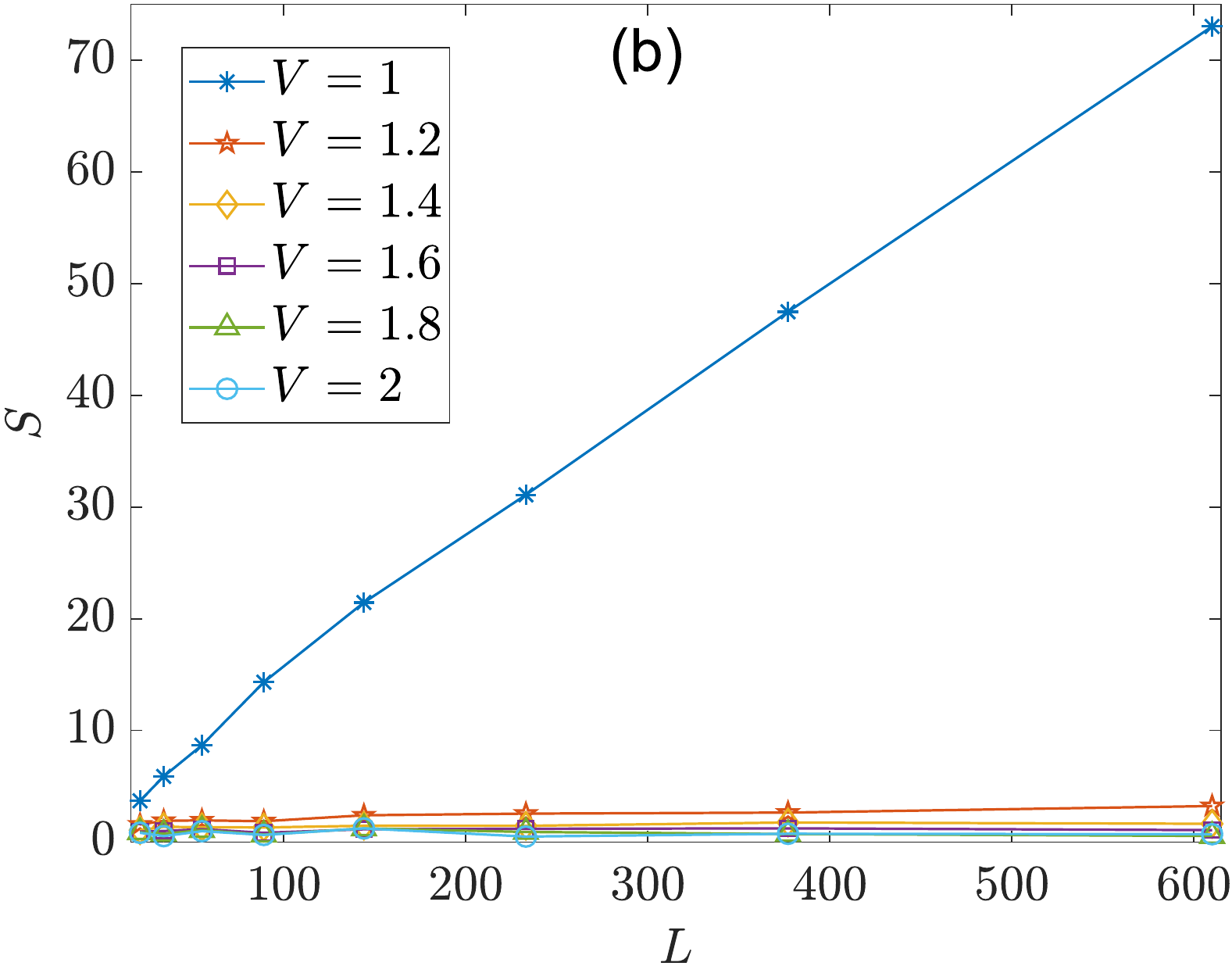}
		\par\end{centering}
	\begin{centering}
		\includegraphics[scale=0.226]{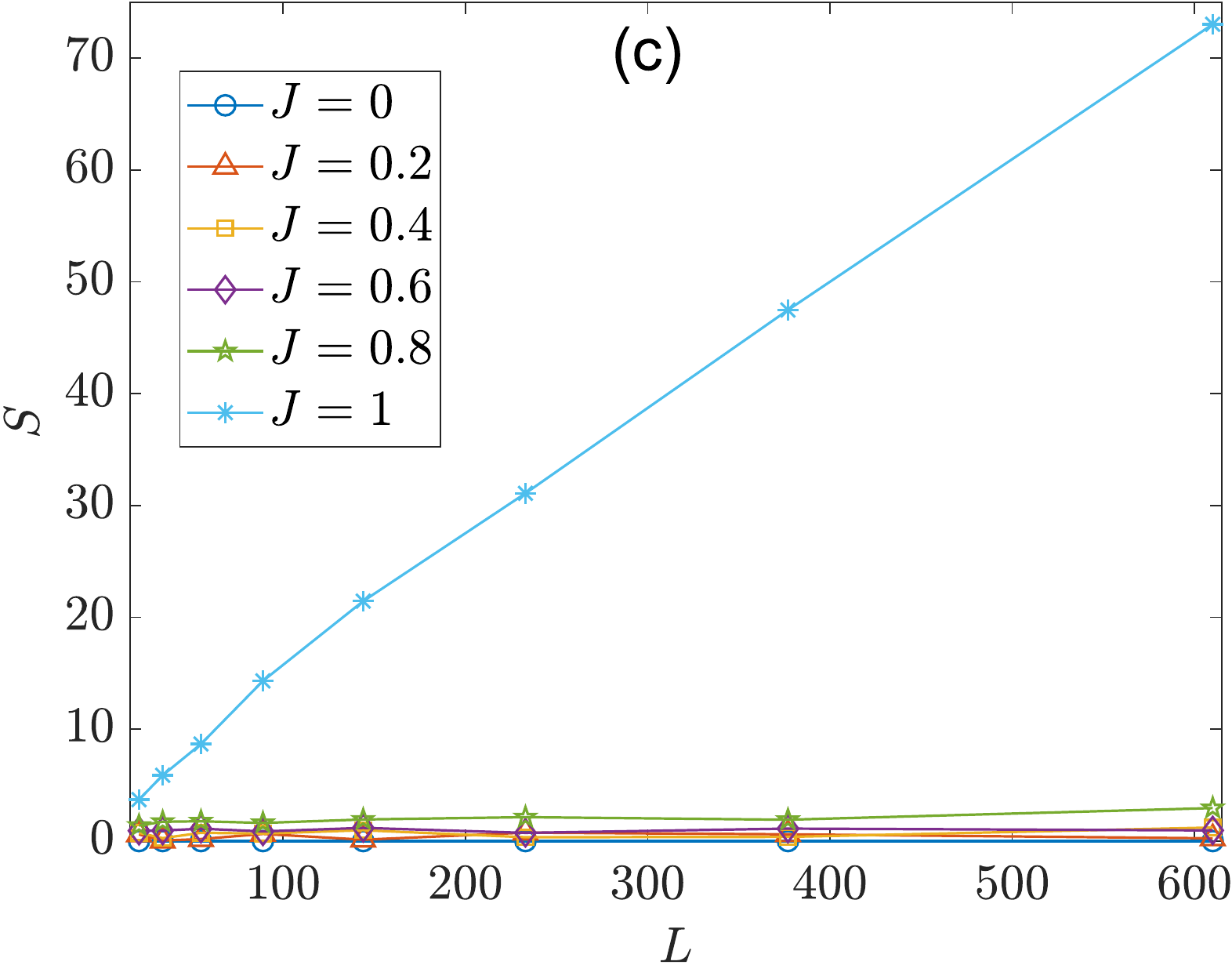}$\,$\includegraphics[scale=0.35]{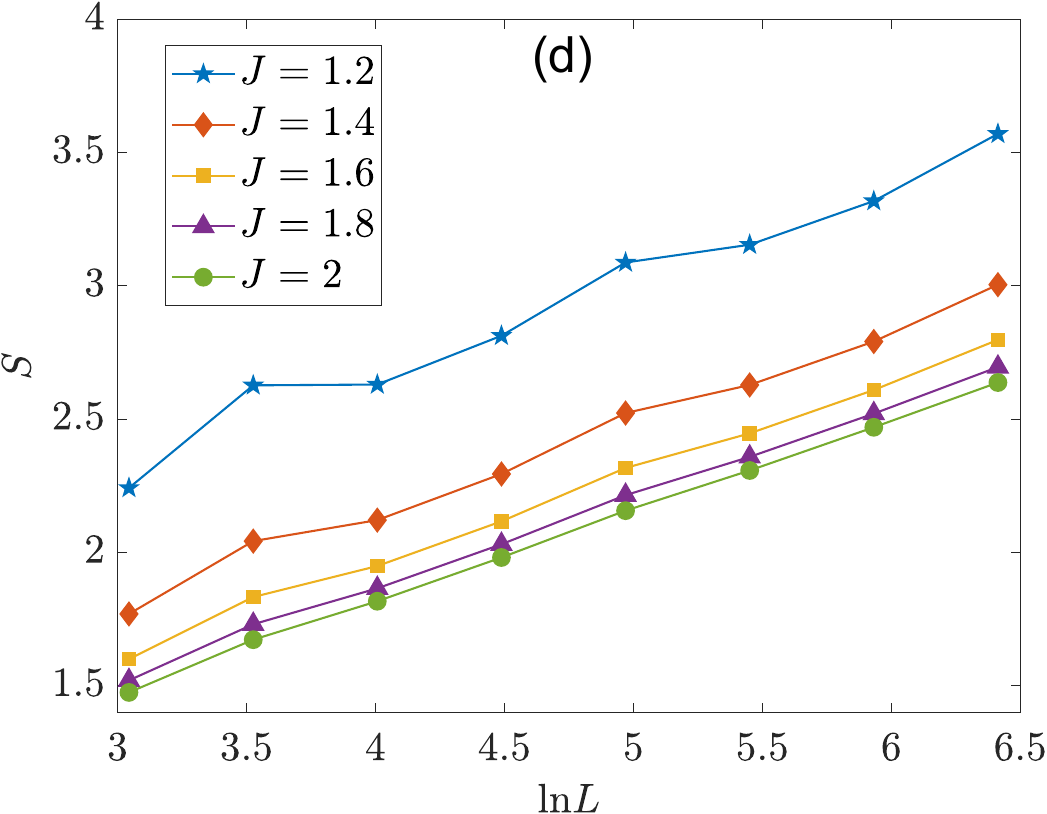}
		\par\end{centering}
	\caption{Steady-state EE of the NHAAH2 versus the system size $L$ at half-filling
		and under equal bipartition. Other system parameters are set as $J=1$
		for (a), (b) and $V=1$ for (c), (d). The time span of the entire
		dynamics is $T=1000$ \cite{Note2}. \label{fig:EEvsL2}}
\end{figure*}

\subsection{NHAAH2\label{subsec:NHAAH2}}

We now explore the entanglement phase transitions in the NHAAH2 [Eq.~(\ref{eq:H2})] by inspecting
the steady-state EE $S(L,l)$ of a subsystem A, where $L$ is the
length of lattice and $l$ is the subsystem size. The initial state
of the system is still at half-filling and described by the wavefunction
$|\Psi_{0}\rangle$ in Eq.~(\ref{eq:Psi0}). Evolving $|\Psi_{0}\rangle$
over a long time duration $T$ from $t=0$, we obtain the EE $S(t)$
at each $t\in[0,T]$ according to Eqs.~(\ref{eq:Psit})--(\ref{eq:St}).
The steady-state $S(L,l)$ is then extracted by averaging $S(t)$
over a time duration $t\in[T',T]$ for an appropriately chosen $1\ll T'<T$.
We could then analyze the scaling behavior of $S(L,l)$ with respect
to the system size $L$ or the subsystem size $l$ at any give sets
of system parameters $(J,V)$.

Similar to the NHAAH1, we first consider a bipartite system with $l=\lfloor L/2\rfloor$
for the NHAAH2. The $L$-dependence of $S(L,L/2)$ for some typical
cases are then obtained and shown in Fig.~\ref{fig:EEvsL2}. We find
that the EE almost does not change with $L$ for $|V|>|J|$, which
suggests that the PT-invariant localized phase of the NHAAH2 is area-law
entangled. At $J=V=1$, we find that up to the leading order $S(L,L/2)\sim gL$,
with the gradient $g\approx0.1$. The same scaling law is found for
other values of $J=V\neq0$, which indicates that the NHAAH2 is volume-law
entangled along the critical lines $J=\pm V$ of the PT-breaking and
localization transitions. Interestingly, we find that up to the leading
order $S(L,L/2)\sim g\ln L$ for the cases with $|V|<|J|$, where
the coefficient $g\approx0.34$. Therefore, the PT-broken extended
phase of the NHAAH2 tends out to be log-law entangled. Such an abnormal
entanglement behavior is clearly distinct from typical scaling laws
of steady-state EE found in other non-Hermitian systems due to non-Hermitian skin effects or
line dissipation gaps \cite{NHEPT04,NHEPT05}. The qualitative change
in the scaling law of steady-state EE from $|V|<|J|$ to $|V|>|J|$
further suggests a log-law to area-law entanglement transition, which
is mediated by a critical volume-law entangled phase along $|V|=|J|$.

\begin{figure*}
	\begin{centering}
		\includegraphics[scale=0.23]{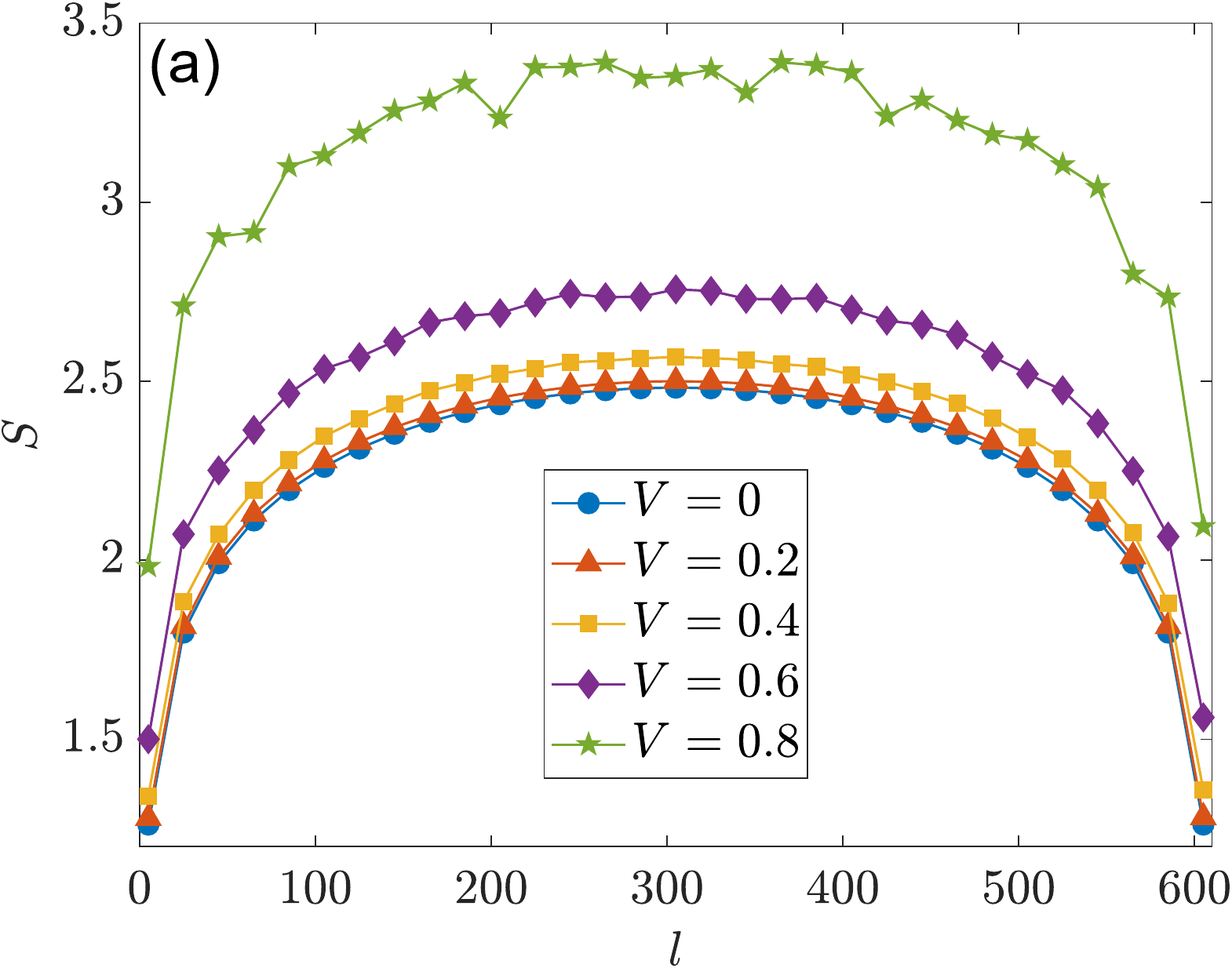}$\,$\includegraphics[scale=0.23]{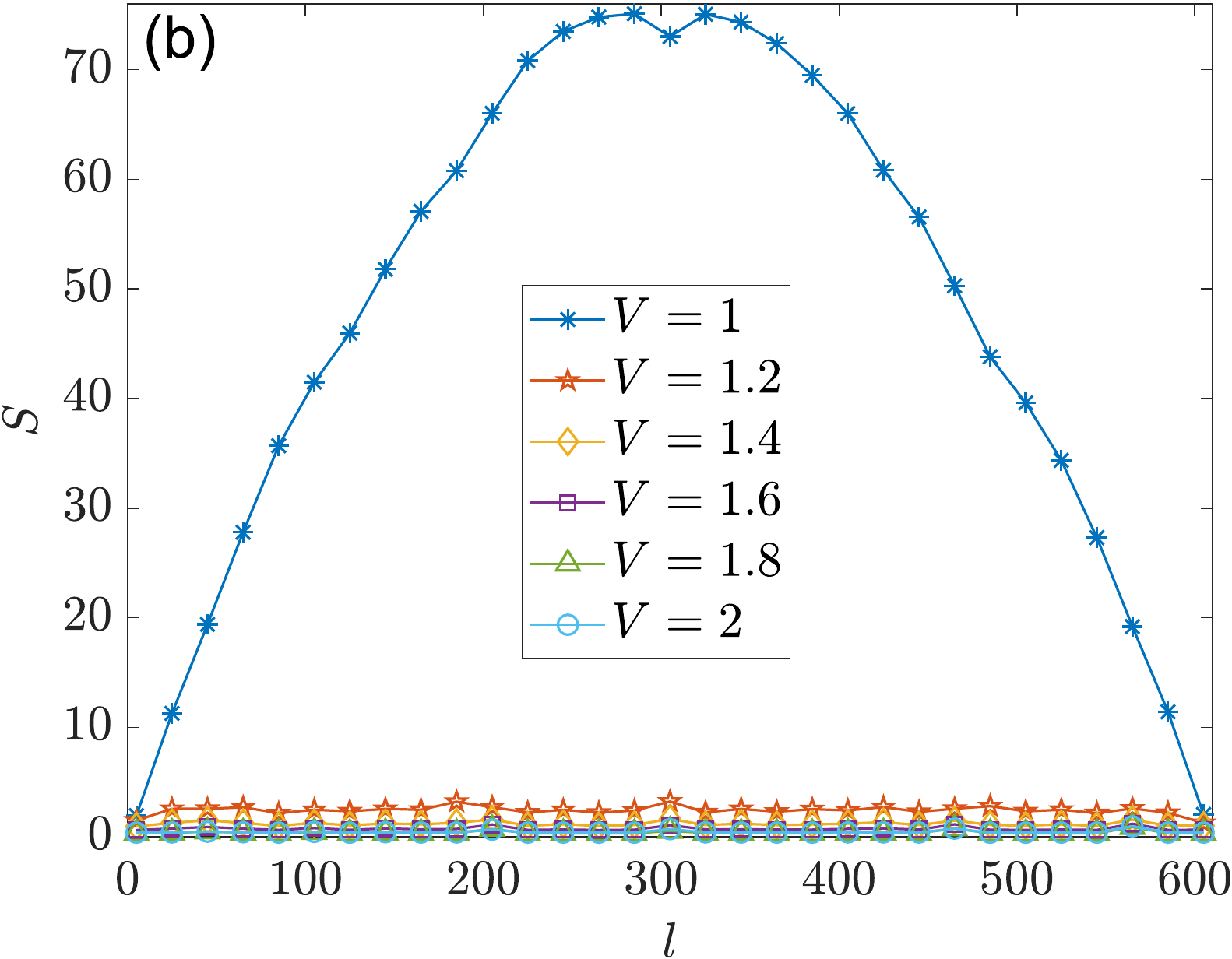}
		\par\end{centering}
	\begin{centering}
		\includegraphics[scale=0.23]{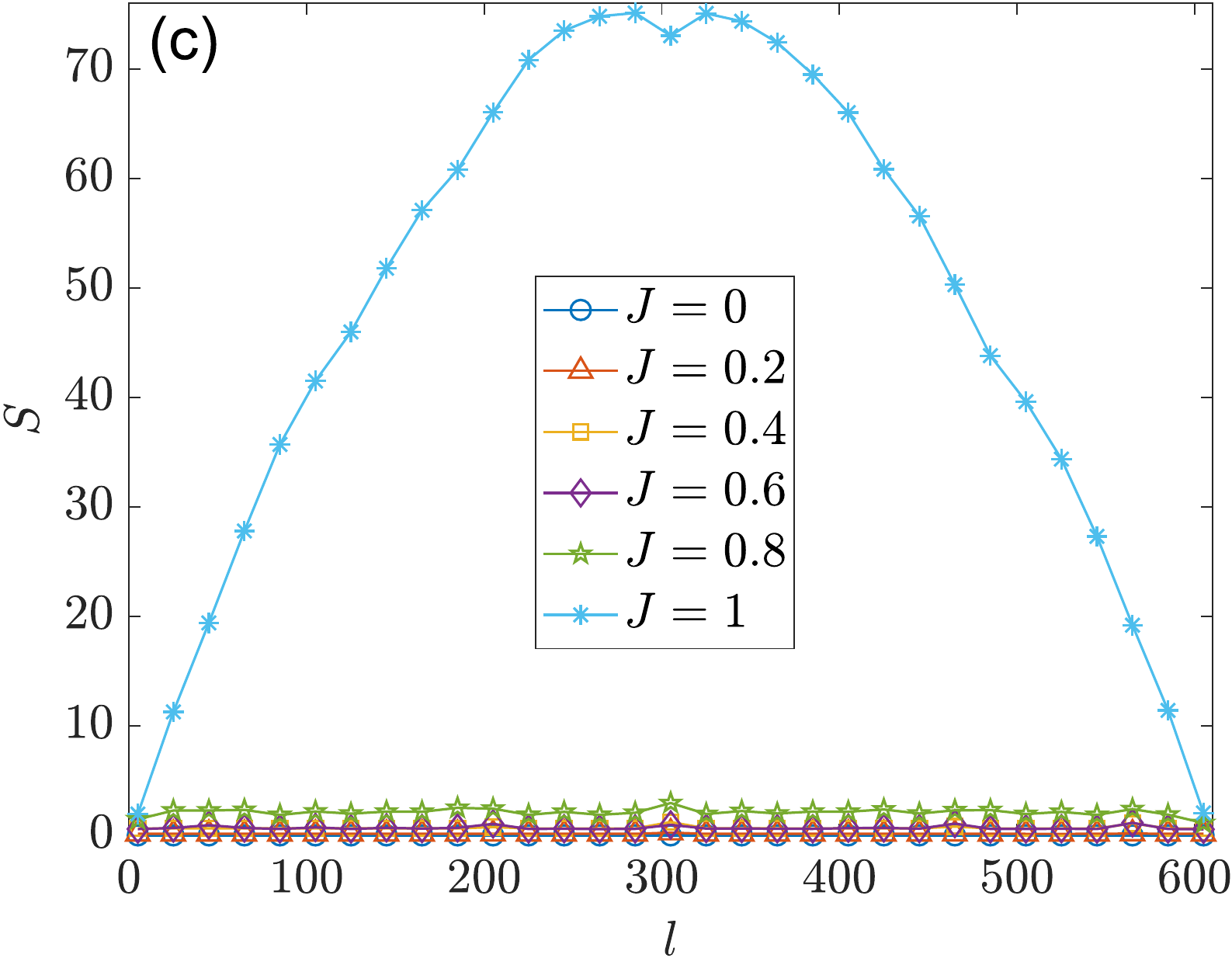}$\,$\includegraphics[scale=0.23]{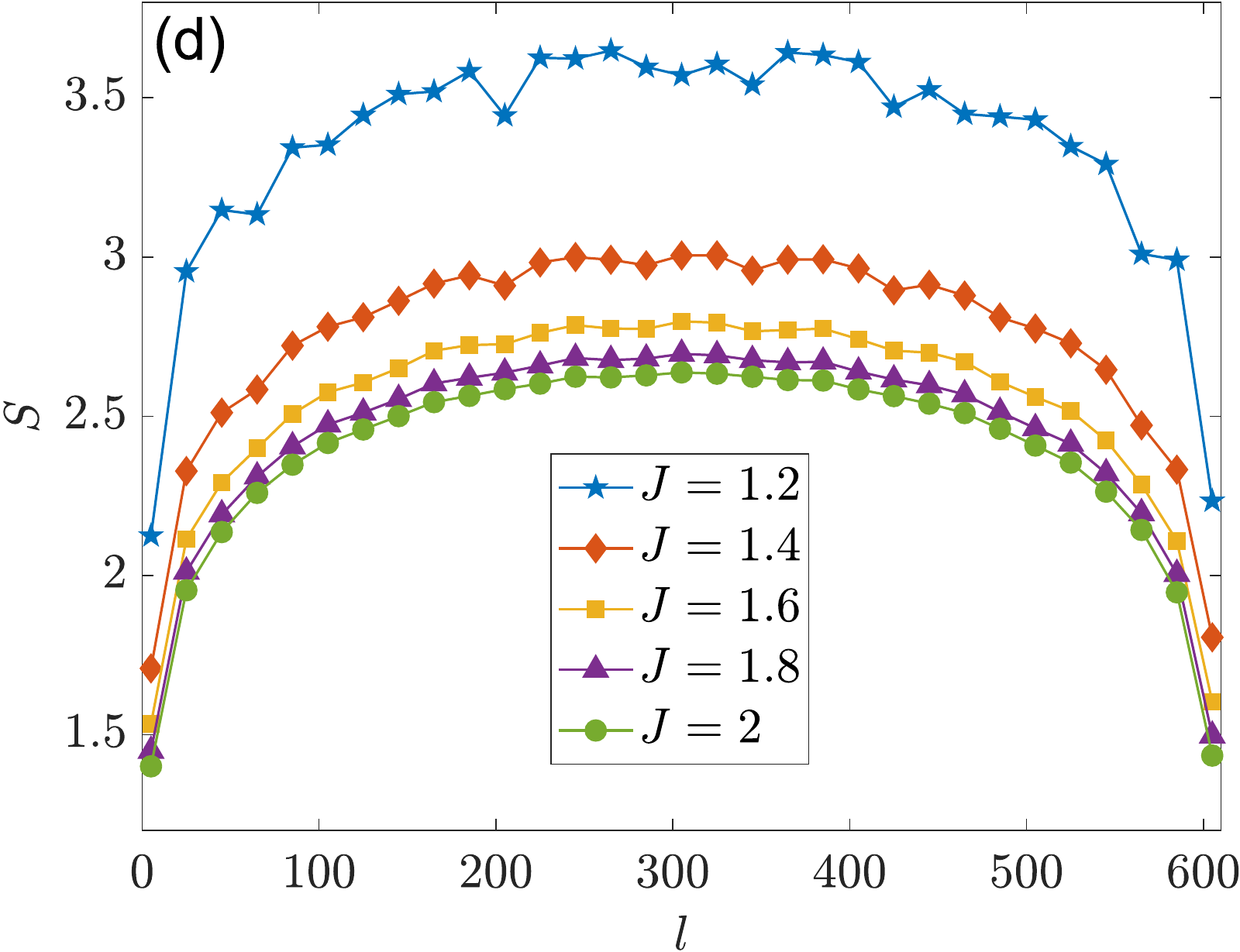}
		\par\end{centering}
	\caption{Steady-state EE of the NHAAH2 vs the subsystem size $l$ at half-filling
		and under a fixed length of lattice $L=610$. Other system parameters
		are $J=1$ for (a), (b) and $V=1$ for (c), (d). The time span
		of the entire dynamics is $T=1000$ \cite{Note2}. \label{fig:EEvsLS2}}
\end{figure*}
To further decode the entanglement transitions in the NHAAH2, we consider
the EE $S(L,l)$ versus the subsystem size $l$ for a fixed $L$,
with typical results at different system parameters shown in 
Fig.~\ref{fig:EEvsLS2}. For the cases with $|V|>|J|$, we find that the
$S(L,l)$ is almost independent of $l$ up to small oscillations,
which is typical for an area-law entangled phase. At $J=V=1$, the
$S(L,l)$ has the shape of the function $A\sin(\pi l/L)+B\ln[\sin(\pi l/L)]+C$
with a small offset at $l=L/2$. Interestingly, our numerics suggest
the following generic form of EE for $|V|<|J|$, i.e.,
\begin{equation}
	S(L,l)\simeq\frac{c}{6}\ln[\sin(\pi l/L)]+S_{0},\label{eq:SLl2}
\end{equation}
where $S_{0}$ is a non-universal constant. 
Away from the transition point $|V|=|J|$, the value of $c$ is found to be $2$ with the numerical error of order $10^{-3}$.
Referring to the typical
form of $S(L,l)$ for a one-dimensional (1D) quantum critical system
\cite{NHEPT04}, Eq.~(\ref{eq:SLl2}) implies a central charge $c=2$
for the PT-broken extended phase of the NHAAH2. The physical origin
of this central charge might be understood from the fact that at half
filling, there are two Fermi points with $E=0$ at $k=\pm\pi/2$ on
the Fermi surface {[}see Eq.~(\ref{eq:H2E}){]}. Each of them makes
a contribution one to the central charge $c$. Compared with the forms
of $S(L,l)$ in Figs.~\ref{fig:EEvsLLS1}(b) and \ref{fig:EEvsLLS1}(d) for the NHAAH1,
we further realize that the NHAAH2 should indeed possess a phase with
unique entanglement nature as described by the scaling relation (\ref{eq:SLl2}).

\begin{figure*}
	\begin{centering}
		\includegraphics[scale=0.35]{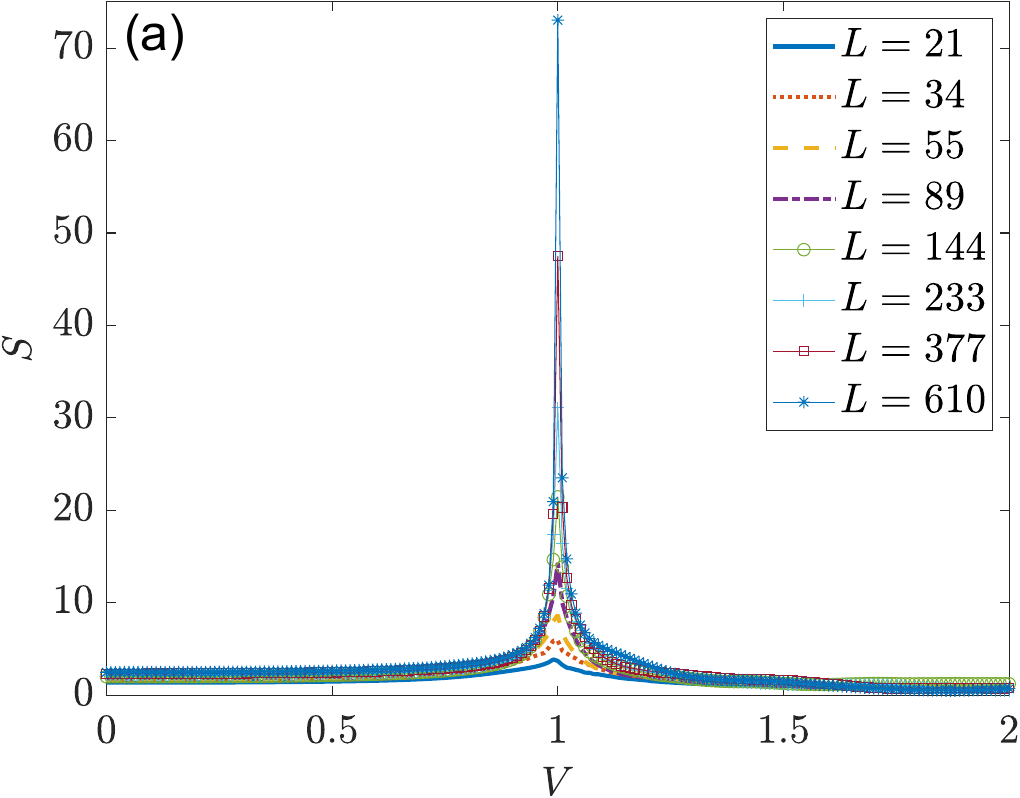}$\,\,$\includegraphics[scale=0.35]{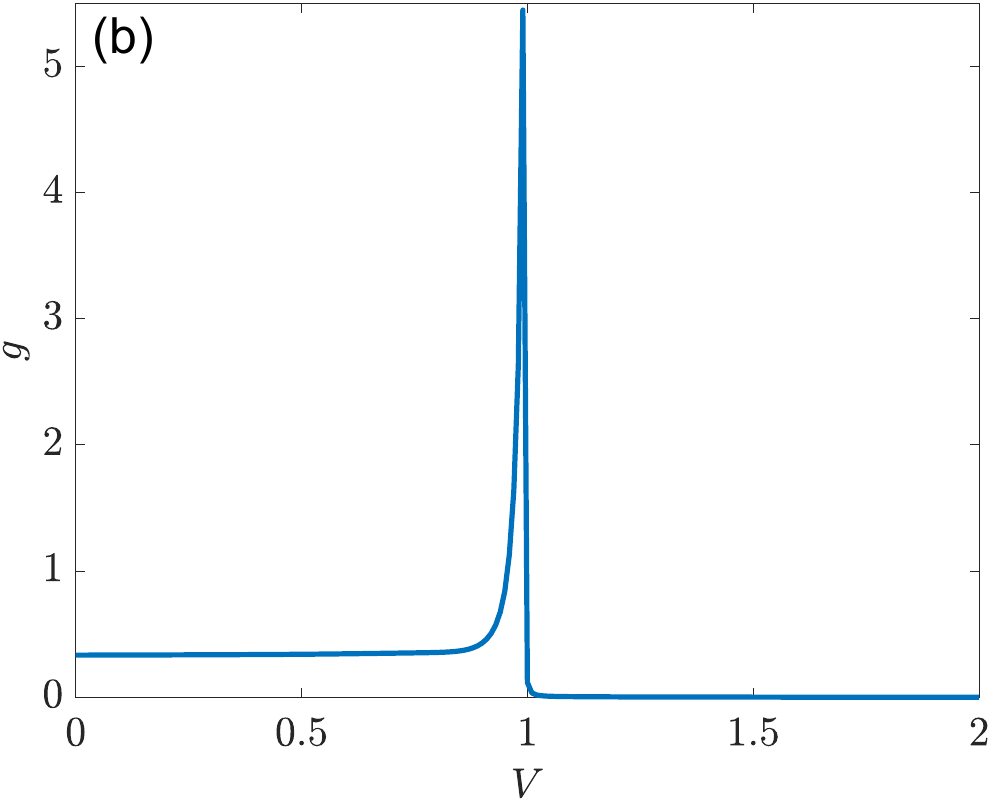}
		\par\end{centering}
	\begin{centering}
		\includegraphics[scale=0.35]{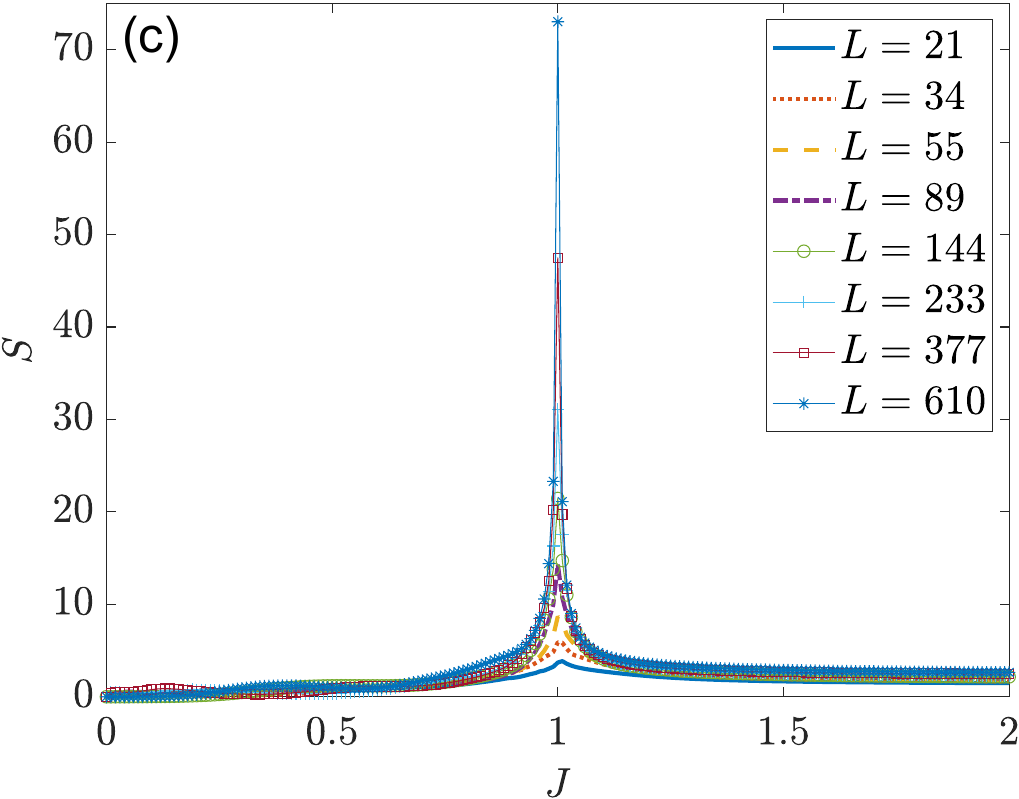}$\,\,$\includegraphics[scale=0.35]{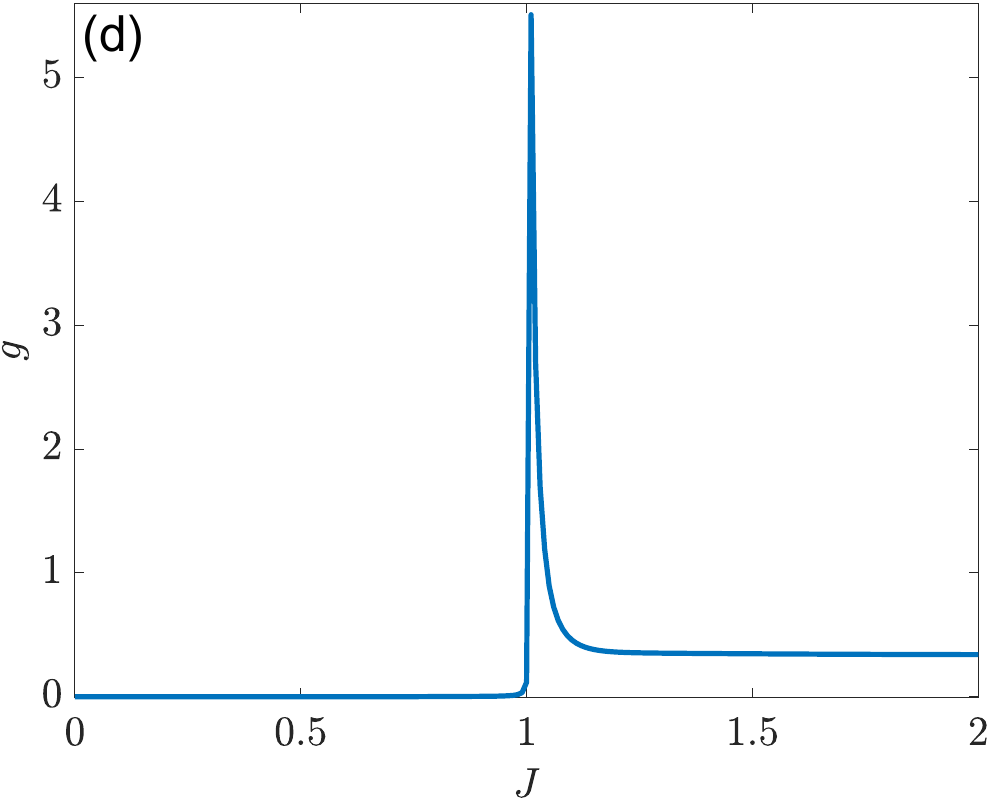}
		\par\end{centering}
	\caption{Bipartite EE of the steady state at half-filling {[}in (a), (c){]}
		and the related gradient $g$ in the scaling law of steady-state EE
		{[}in (b) and (d){]} for the NHAAH2. Other system parameters are set
		as $J=1$ for (a), (b) and $V=1$ for (c), (d). The time span of the
		entire evolution is $T=1000$ \cite{Note2}. The values of $g$ are obtained from
		the linear fitting $S\sim gL+s_{0}$ ($S\sim g\ln L+s_{0}$) of EE
		versus the system size $L$ for $J\protect\leq V$ ($J>V$) in (b)
		and (d). \label{fig:EEvsJV2}}
\end{figure*}
Combining the information obtained from the scaling properties of
EE with respect the system size, we are now ready to reveal the entanglement
phase transitions in the NHAAH2. In Figs.~\ref{fig:EEvsJV2}(a) and
\ref{fig:EEvsJV2}(c), we present the steady-state EE versus $V$ and $J$ for different
system sizes. A clear peak can be identified at $J=V$, whose height
increases monotonically with the increase of the lattice size $L$.
The presence of such a sharp peak in $S(L,L/2)$ clearly hints at
the occurrence of a entanglement transition at $J=V$. In Figs.~\ref{fig:EEvsJV2}(b)
and \ref{fig:EEvsJV2}(d), we use the relations $S\sim gL+s_{0}$ and $S\sim g\ln L+s_{0}$
to fit the data at different $L$ for $|J|\leq|V|$ and $|J|>|V|$,
respectively. The results suggest that the scaling form of EE could
undergo a discontinuous change from a log-law ($|V|<|J|$) with a
finite $g$ in $S\sim g\ln L+s_{0}$ to an area-law with $g\simeq0$
in the linear fitting $S\sim gL+s_{0}$ ($|V|>|J|$). There is thus
an entanglement phase transition at $|V|=|J|$ accompanying the PT
and localization transitions in the NHAAH2.

To have a more balanced comparison between the scaling laws in different parameter regions, we could assume a fitting function $S(L,L/2)\sim g{\ln}L+g'L+s_0$. Our numerical calculations then suggest that $g\simeq0.34$ and $g'\simeq0$ in the region $|V|<|J|$, while $g\simeq0$ and $g'\simeq0$ in the region $|V|>|J|$. These two regions thus correspond to log-law and area-law entangled phases. Right at $|V|=|J|$, we find $g\simeq2.4$ and $g'\simeq0.1$, and the volume-law scaling dominants with the increase of $L$.

\begin{figure}
	\begin{centering}
		\includegraphics[scale=0.48]{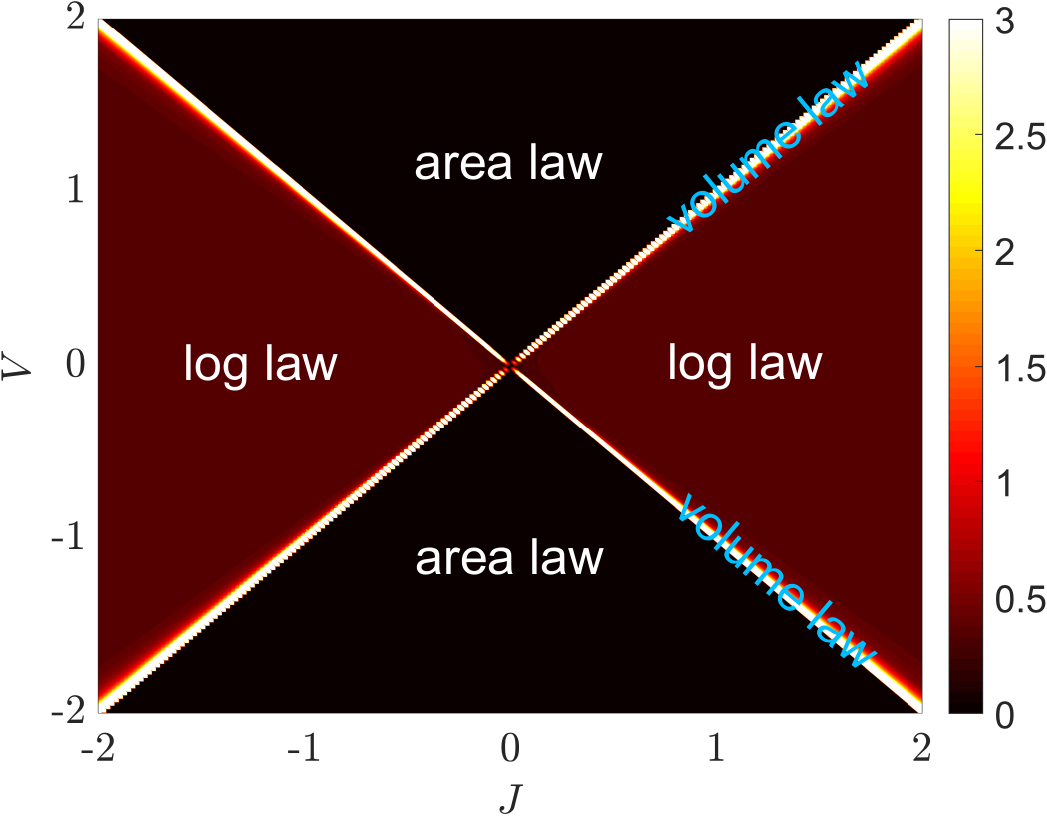}
		\par\end{centering}
	\caption{Entanglement phase diagram of the NHAAH2. Different colors correspond
		to different values of the gradient $g$ extracted from the fitting
		$S\sim gL+s_{0}$ ($S\sim g\ln L+s_{0}$) of steady-state EE versus
		the system size $L$ for $|V|\geq|J|$ ($|V|<|J|$). \label{fig:EEPhsDiag2}}
\end{figure}

\begin{table*}
	\begin{centering}
		\begin{tabular}{|c|c|c|c|}
			\hline 
			NHAAH2 & $|V|<|J|$ & $|V|=|J|$ & $|V|>|J|$\tabularnewline
			\hline 
			\hline 
			Energy spectrum & complex & PT transition & real\tabularnewline
			\hline 
			Eigenstates & extended & localization transition & localized\tabularnewline
			\hline 
			Steady-state EE & log-law & volume-law & area-law\tabularnewline
			\hline 
		\end{tabular}
		\par\end{centering}
	\caption{Summary of main results for the quasicrystal NHAAH2 \ref{eq:H2}. The complex-spectrum
		(PT-broken), extended phase is log-law entangled. The real-spectrum
		(PT-invariant), localized phase is area-law entangled. The PT, localization
		and entanglement transitions happen all together at $|V|=|J|$,
		where the steady-state EE follows the volume-law as in NHAAH1 [see also Figs.~\ref{fig:E-IPR}(c), \ref{fig:E-IPR}(d) and \ref{fig:EEPhsDiag2}].}\label{Tab2}
\end{table*}

Finally, we establish the entanglement phase diagram of the NHAAH2
by extracting the scaling laws of steady-state EE versus the system
size $L$ for a half-filled and bipartite lattice under the PBC, as shown
in Fig.~\ref{fig:EEPhsDiag2}. We observe that the EE indeed satisfies
an area law {[}$S(L,L/2)\sim L^0${]} in the PT-invariant localized
phase ($|J|<|V|$), and fulfills an anomalous log-law scaling {[}$S(L,L/2)\propto\ln L${]}
in the PT-broken extended phase ($|J|>|V|$). Along the phase boundary
($|J|=|V|$), the EE shows a volume-law critical scaling behavior
{[}$S(L,L/2)\propto L${]}, which is similar to the NHAAH1. Besides
that, the entangled phases and entanglement transitions in the NHAAH2
are rather different from those appeared in NHAAH1.
A summary of the key features of NHAAH2 is given in Table \ref{Tab2}.

A possible reason behind these differences is as follows. In the PT-broken
extended phase of NHAAH2 ($|J|>|V|$), the asymmetric hopping overcomes
the block of quasiperiodic disorder and allows the spreading of quantum
information across the system, yielding the tendency of forming an
extensively entangled phase. However, the spectrum of the system in
the PT-broken phase possesses a point gap on the complex energy plane
at $E=0$ {[}see Eq.~(\ref{eq:H2E}){]}. The presence of such a dissipation
gap tends to suppress the quantum information spreading and prefers
an area-law scaling for the steady-state EE. The competition between
these two opposite tendencies ends up with a compromise, as reflected
by the log-law entangled phase in Fig.~\ref{fig:EEPhsDiag2}. In the
PT-invariant localized phase of NHAAH2 ($|J|<|V|$), the disorder
is strong enough to prevents the information spreading and stabilizes
the system in an area-law entangled phase, even though the energy
spectrum is fully real {[}see Eq.~(\ref{eq:H2E}){]}. The very different
entanglement dynamics in our mutually dual NHQC models could thus
be understood. The clear differences between the entanglement transitions
discovered here and some typical situations encountered in previous
studies \cite{NHEPT01,NHEPT04,NHEPT05} further highlight the interesting
role played by disorder in non-Hermitian systems from a quantum information
perspective.

\section{Conclusion and discussion\label{sec:Sum}}
In this work, we revealed entanglement phase transitions in representative
1D NHQCs. In a system with onsite gain and loss, a volume-law to area-law
transition in the steady-state EE was found to go hand-in-hand with
PT-breaking and localization transitions induced by non-Hermitian
quasiperiodic potentials. In a system with nonreciprocal hopping,
the steady-state EE instead showcased an area-law to log-law entanglement
transition with the increase of the hopping asymmetry, which was mediated
by a critical entangling phase whose EE followed a volume-law scaling
versus the system size. This transition also went hand-in-hand with
PT-breaking and delocalization transitions due to the interplay between
hopping nonreciprocity and spatial quasiperiodicity. Even though the
two considered models can be viewed as dual to each other, they exhibited
rather different entanglement dynamics except at critical points, which were demonstrated in
detail by our numerical analysis of their EE scaling laws and entanglement
phase diagrams. Our findings thus unveiled the richness of entanglement
phases and transitions in non-Hermitian disordered systems, which
may find applications in quantum error correction and quantum information
storage against decoherence.

As we focused on phases and entanglement dynamics of the bulk of NHQCs, the PBC was taken throughout our calculations. A consistent framework regarding PT transitions, localization transitions and entanglement transitions were then established for our ``minimal'' NHQC models under PBC, and rich patterns of entanglement transitions were identified. Under open boundary conditions, there could be edge states in our models, whose numbers are much smaller than the bulk states. The NHAAH2 would further show non-Hermitian skin effects. A complete treatment of their interplay with entanglement transitions under different boundary conditions would thus be an interesting direction of future research.

In Figs.~\ref{fig:EEvsLLS1} and \ref{fig:EEvsLS2}, some asymmetries are observed in $S(L,l)$ vs $l$ with $L=610$ when the system parameters are approaching the phase boundaries ($|J|=|V|$ for both models). One possible origin of these asymmetries is the instability of numerical calculations around the critical points of phase transitions. Another possible source is that at the critical point of localization transition, the quasicrystal may show multifractal properties, and correction terms other than the volume-law or log-law may appear in the subsystem-size scaling of EE, even though the volume-law or log-law behavior still dominants. Our numerical resolutions could not figure out all these correction terms at present. An in-depth analysis about the critical properties and universality classes of entanglement phase transitions in NHQCs is thus necessary in future studies.

Our model NHAAH1 [Eq.~(\ref{eq:H1})] possesses onsite gain and loss. In theory, it might be viewed as the no-click limit of a monitored AAH model. The gain and loss in the system may then be understood as imaginary chemical potentials induced by measurement backactions \cite{EPT52,NHEPT05}. In practice, cold atom systems \cite{NHCdAtm01,NHCdAtm02,NHCdAtm03,NHCdAtm04,NHCdAtm05} could be considered as candidates to realize our models. For fermions, one may introduce state-selective atom loss by using a near-resonant laser beam to kick atoms out of a trap \cite{NHCdAtm04}. The negative imaginary part of onsite potential in our NHAAH1 then describes the loss rate. Realizing atom gain for fermions is more challenging. One may instead add a uniform background loss $-i\gamma\sum_n{\hat c}^\dagger_n{\hat c}_n$ with $\gamma>0$ to our NHAAH1 and let $\gamma>|V|$. In this case, there is no gain in the system, yet the Hamiltonian loses its PT-symmetry in strict sense. Nevertheless, we can still find the spectrum transformation from a line segment (with $|V|<|J|$) to an ellipse (with $|V|>|J|$), which is now centered at $(0,-i\gamma)$ on the complex plane. The particle dynamics and EE dynamics are not affected by such a uniform background loss according to Eqs.~(\ref{eq:Psit})--(\ref{eq:St}). Additionally, the unidirectional hopping of our NHAAH2 might be realized by implementing asymmetric quantum walks of cold atoms in momentum space \cite{NHCdAtm05}, which is not sensitive to particle statistics. Putting together, our non-Hermitian fermionic models should be physically realizable in near-term experiments.

Although our results are obtained by investigating two ``minimal''
NHQC models, we expect to find similar patterns of entanglement phase
transitions in other 1D NHQCs with simultaneous PT and localization
transitions, such as those considered in Refs.~\cite{NHQC04,NHQC05,NHQCExp1}.
In more general situations, the extended and localized phases of an NHQC
could be separated by a critical phase, in which extended and localized
eigenstates coexist and are separated by mobility edges. The entanglement
transition in NHQCs with mobility edges thus constitutes another interesting
direction of future research. Besides, much less is known regarding
entanglement transitions in non-Hermitian disordered systems beyond
one spatial dimension, with uncorrelated disorder \cite{NHAI01,NHAI02,NHAI03}
and with many-body interactions \cite{NHMBL01,NHMBL02}. Concrete experimental signatures
of entanglement phase transitions in non-Hermitian systems also deserve
more thorough considerations.

\textit{Note added}: Before the submission of this work, we realized a new preprint \cite{ZLarXiv}, which also explored entanglement phase transitions in NHQCs with a focus on the interply between disorder and non-Hermitian skin effects.

%\vspace{0.5cm}
\begin{acknowledgments}
	This work is supported by the National Natural Science Foundation of China (Grant Nos.~12275260 and 11905211), the Fundamental Research Funds for the Central Universities (Grant No.~202364008), and the Young Talents Project of Ocean University of China.
\end{acknowledgments}

%\appendix


\begin{thebibliography}{99}
\bibitem{Note0x} Eqs.~(\ref{eq:H1E}) and (\ref{eq:H2E}) are conjectured forms of energy spectrum based on extended numerical calculations \cite{NHQC02,NHQC06,NHQC13}. In the limit of large system size $L\rightarrow\infty$, one may view $k$ as a synthetic quantum number of the spectrum.
\end{thebibliography}
\end{document}